\documentclass[twocolumn,showpacs,pra]{revtex4-1}

\usepackage{amsmath,amssymb,graphicx}

\begin{document}

\title{Coherent light in intense spatio-spectral twin beams}

\author{Jan Pe\v{r}ina Jr.}
\email{jan.perina.jr@upol.cz}
\affiliation{RCPTM, Joint Laboratory
of Optics of Palack\'{y} University and Institute of Physics of
Academy of Sciences of the Czech Republic, Faculty of Science,
Palack\'{y} University, 17. listopadu 12, 771~46 Olomouc, Czech
Republic}

\begin{abstract}
Intense spatio-spectral twin beams generated in the regime with
pump depletion are analyzed applying a suggested quantum model
that treats the signal, idler and pump fields in the same way. The
model assumes the signal and idler fields in the form of the
generalized superposition of signal and noise and reveals nonzero
signal coherent components in both fields, contrary to the models
developed earlier. The influence of coherent components on the
properties of intense twin beams is elucidated. The interference
pattern formed in the process of sum-frequency generation and that
of the Hong-Ou-Mandel interferometer are shown to be able to
experimentally confirm the presence of coherent components.
\end{abstract}



\maketitle

\section{Introduction}

Optical parametric processes \cite{Boyd2003} are the most
frequently studied nonlinear optical processes due to their
relatively high efficiency compared to other nonlinear processes
\cite{Butcher1991}. They involve in general the whole family of
optical processes describing the interaction of three fields:
sum-frequency generation, difference-frequency generation,
parametric amplification and oscillation, or frequency up- and
down-conversion \cite{Boyd2003}. They include, as the degenerate
case, the process of second-harmonic generation --- the first
observed nonlinear optical process \cite{Franken1961}. Optical
parametric processes have been investigated for various types of
optical fields including strong coherent and chaotic fields
\cite{Boyd2003} as well as highly nonclassical quantum fields with
their intensities at the single photon level \cite{Mollow1967,
Mollow1967a,Perina1991}. Fields composed of photon pairs in the
Fock states with one photon localized in the signal field and the
other in the idler field occupy a prominent position among such
quantum fields due to their highly unusual properties
\cite{Mandel1995}. When the multi-mode signal and idler fields are
considered, such states are entangled in various degrees of
freedom, as a consequence of specific conditions and rules
governing the photon-pair emission \cite{Ghosh1986}. In this case,
photon pairs emerge in the process of spontaneous parametric
down-conversion (PDC). This process has also its intense variant,
stimulated parametric down-conversion, that provides the so-called
twin beams (TWB) composed of many photon pairs. Whereas some
nonclassical properties of individual photon pairs are concealed
in such more intense TWBs, other interesting and attractive
properties remain. Namely, these are the sub-shot-noise intensity
correlations between the signal and idler fields
\cite{Nabors1990,Tapster1991,Jedrkiewicz2004,Bondani2007,Blanchet2008,Brida2009a}
and 'tight' spectral and spatial intensity correlations
\cite{Kolobov1989,Kolobov1989a,Brida2010,Machulka2014,Haderka2015}.
Moreover, more intense TWBs belong to macroscopic fields, i.e.
fields detectable by classical detectors.

For this reason, intense TWBs have attracted considerable
attention in the last twenty years
\cite{Gatti2003,Brambilla2004,Brambilla2010,Caspani2010,Christ2011,Stobinska2012,Perez2014,Chekhova2015,Sharapova2015,Cavanna2016}.
Their spectral as well as spatial intensity correlation functions
have been determined both theoretically and experimentally.
Sub-shot-noise intensity correlations have even been
experimentally exploited in quantum imaging \cite{Genovese2016}
that reveals an image via ideally-noiseless sub-shot-noise
intensity correlations \cite{Brida2010a}. They have also found
their applications in spectroscopy \cite{SoutoRibeiro1997} and
quantum interferometry \cite{RuoBerchera2015}, where they
qualitatively improve the experimental precision. Attention has
also been payed to photocount statistics in the signal (or idler)
field that was identified as multi-mode chaotic (thermal)
\cite{Allevi2014}, owing to the spontaneous emission of photon
pairs at the initial stage of the TWB evolution
\cite{Haderka2005a,PerinaJr2012}. This is a bit surprising as the
stimulated PDC plays an important role in the creation of more
intense TWBs. In theory, the multi-mode chaotic statistics is
found in the models that assume sufficiently intense pump fields
not depleted during the nonlinear interaction
\cite{PerinaJr2015a}.

However, the measurement of intensity correlation functions of
TWBs such intense that the pump field is depleted during the
interaction has revealed their unexpected behavior
\cite{Allevi2014a,Allevi2014b,Allevi2015,Allevi2015a}. Whereas the
models with an un-depleted pump field predict broadening of
spectral and spatial intensity auto- and cross-correlation
functions of TWBs \cite{Gatti2003,Brambilla2010,PerinaJr2015a},
narrowing of these correlation functions has been observed for
sufficiently intense pump fields \cite{Allevi2014a}. This has
pointed out at the complicated internal dynamics of TWBs during
their creation \cite{PerinaJr2016a}. Applying the Schmidt modes
for the signal and idler fields in the analysis of the nonlinear
interaction, the strong flow of energy between the individual pump
modes and their paired signal and idler counterparts (Schmidt
modes) has been revealed \cite{PerinaJr2016}. This energy flow is
so strong that the most efficiently populated paired signal and
idler modes reach their maximal photon numbers for the pump powers
only slightly greater than the threshold pump power at which the
coherence maxima are observed. For even higher powers, these modes
attain their maximal photon numbers somewhere inside the crystal
and then loose their energy in favor of their pump modes. Typical
maximal photon numbers reached under real experimental conditions
are of the order $ 10^6 - 10^8 $ photons per mode. This shows that
the stimulated emission has to play the dominant role in the
evolution of individual modes.

The observed coherence maxima have been successfully explained by
a model based on the generalized parametric approximation
\cite{PerinaJr2016}. In this approximation, the genuine quantum
momentum operator
$$
  i\hbar K \hat{a}_{\rm p}(z) \hat{a}_{\rm s}^{\dagger}(z)
   \hat{a}_{\rm i}^{\dagger}(z) + {\rm h.c.},
$$
written in the position-dependent creation [$ \hat{a}^\dagger(z)
$] and annihilation [$ \hat{a}(z) $] operators of the pump ($ \rm
p $), signal ($ \rm s $) and idler ($ \rm i $) modes, is
substituted by the following momentum operator
$$
  i\hbar K A_{\rm p}(z) \hat{a}_{\rm s}^{\dagger}(z)
   \hat{a}_{\rm i}^{\dagger}(z) + {\rm h.c.}
$$
that assumes a classical position-dependent pump-field amplitude $
A_{\rm p}(z) $. This approximative form of the momentum operator
assumes that there occurs no classical (coherent) amplitude in the
signal and idler fields, even when the most of the energy is moved
from the pump mode into the corresponding signal and idler modes.
In our opinion, this does not accord with the important role of
stimulated PDC in the process and the above momentum operator
should be extended to account for this. Here, we suggest the
following natural extension of the above momentum operator that
treats all three interacting fields in the same way and uses the
classical position-dependent signal- [$ A_{\rm s}(z) $] and idler-
[$ A_{\rm i}(z) $] field amplitudes:
\begin{eqnarray*}
  & i\hbar K \biggl[ A_{\rm p}(z)\hat{a}_{\rm s}^{\dagger}(z)
   \hat{a}_{\rm i}^{\dagger}(z) + i\hbar K A_{\rm s}^*(z) \hat{a}_{\rm p}(z)
   \hat{a}_{\rm i}^{\dagger}(z) & \\
  & + i\hbar K A_{\rm i}^*(z) \hat{a}_{\rm p}(z)
   \hat{a}_{\rm s}^{\dagger}(z) \biggr] + {\rm h.c.}&
\end{eqnarray*}
Our analysis below shows that, indeed, this momentum operator
provides coherent components in the signal and idler fields,
together with the usual chaotic contributions. Moreover, these
coherent components influence only weakly the coherence properties
of TWBs compared to the successful model based on the generalized
parametric approximation \cite{PerinaJr2016}. We note that the
transition of the chaotic behavior for low field intensities
towards the coherent one for greater field intensities above the
threshold of an optical parametric oscillator has been analyzed in
terms of the photon-number statistics in \cite{Graham1968a}, where
the Van der Pol equation for a down-converted field has been
derived.

To distinguish the predictions of both models, we analyze the
behavior of intense TWBs in the process of sum-frequency
generation and also in the Hong-Ou-Mandel interferometer. The
model with coherent components predicts the suppression of a
narrow peak in the interference pattern of sum-frequency
generation. On the other hand, it predicts a small peak in the
central part of a broad Hong-Ou-Mandel interference dip. Contrary
to this, the model based on the generalized parametric
approximation suggests a narrow peak on the top of a broad
sum-frequency intensity peak as well as a narrow dip formed in the
central part of the broad Hong-Ou-Mandel interference dip. The
presence of the coherent components in real multi-mode TWBs means
that the observed spectral and spatial speckle grains have both
chaotic and coherent components. We note that no common coherent
field covering the whole TWB is predicted. Also, contrary to the
usual model of multi-mode chaotic TWBs the model with the coherent
components predicts the loss of sub-shot-noise intensity
correlations with the increasing TWB intensity, in agreement with
the experimental observations \cite{Jedrkiewicz2004,Bondani2007}.

The paper is organized as follows. In Sec.~II, the quantum model
of a TWB is introduced and its dynamics is found. Sec.~III is
devoted to the statistical properties of individual modes'
triplets composed of one signal, one idler and one pump mode.
Statistical properties of the whole TWB are described in Sec.~IV.
The behavior of TWBs in the process of sum-frequency generation
and in the Hong-Ou-Mandel interferometer is analyzed in Sec.~V.
Discussion of the behavior of individual modes' triplets for a
typical experimental configuration of PDC is given in Sec.~VI. The
behavior of the whole TWB in the same configuration is analyzed in
Sec.~VII. TWBs participating in the process of sum-frequency
generation and propagating in the Hong-Ou-Mandel interferometer
are investigated in Sec.~VIII. Conclusions are drawn in Sec.~IX.

\section{Quantum model of twin beams with coherent components}

We describe PDC by an approximate momentum operator $ \hat{G}_{\rm
int} $ that is constructed upon mutually independent modes'
triplets containing one mode from the signal, idler and pump
fields. The signal and idler orthonormal modes are revealed by the
Schmidt decomposition of a two-photon amplitude describing photon
pairs at the single-photon level
\cite{Law2000,Law2004,Mikhailova2008,Fedorov2014,PerinaJr2015}. A
pump mode associated with a given signal and idler mode pair is
determined by 'matching' all three modes in the nonlinear
interaction (see below). The momentum operator $ \hat{G}_{\rm int}
$ appropriate for the radially symmetric geometry is written in
the interaction representation as follows
\cite{PerinaJr2015,PerinaJr2015a,Mandel1995,Perina1991}:
\begin{eqnarray}     
 \hat{G}_{\rm int}^{\rm av}(z) &=& i\hbar
  \sum_{m=-\infty}^{^\infty} \sum_{l,q=0}^{\infty}
    K_{mlq} \hat{a}_{{\rm p},mlq}(z) \hat{a}_{{\rm s},mlq}^{\dagger}(z)
   \hat{a}_{{\rm i},mlq}^{\dagger}(z)\nonumber \\
 & & \mbox{}  + {\rm h.c.}
\label{1}
\end{eqnarray}
Symbol $ \hat{a}_{b,mlq} $ ($ \hat{a}^\dagger_{b,mlq} $) in
Eq.~(\ref{1}) denotes an annihilation (creation) operator of a
photon in field $ b $ in the spatio-spectral mode indexed by $ mlq
$. Index $ m $ corresponds to the azimuthal transverse angle,
index $ l $ to the radial transverse direction and index $ q $ is
used in the frequency domain. Nonlinear coupling constants $
K_{mlq} $ differ from mode to mode as a consequence of the varying
overlap of the normalized signal ($ f_{{\rm s},q} $), idler ($
f_{{\rm i},q} $) and pump ($ f_{{\rm p},q} $) spectral modes.
Defining the non-normalized spectral pump modes $ f_{{\rm
p},q}^{({\rm n})} $ along the formula
\begin{equation}   
 f_{{\rm p},q}^{({\rm n})}(\omega_{\rm p}) = \int_{-\infty}^{\infty} d\omega_{\rm s}
 f_{{\rm s},q}(\omega_{\rm s}) f_{{\rm i},q}(\omega_{\rm p}-\omega_{\rm s}),
\label{2}
\end{equation}
the coupling constant $ K_{mlq} $ is determined as $ \tilde{K}
\kappa_q^\parallel $, where $ \kappa_q^\parallel \equiv \sqrt{
\int d\omega_{\rm p} |f_{{\rm p},q}^{({\rm n})}(\omega_{\rm p})|^2
} $ is given by the norm of the non-normalized pump mode $ f_{{\rm
p},q}^{({\rm n})} $ and $ \tilde{K} $ stands for a common coupling
constant. The modes $ f_{{\rm p},q}^{({\rm n})} $ are in general
neither normalized nor orthogonal. We note that the pump modes $
f_{{\rm p},q}^{({\rm n})} $ are properly normalized only for cw
interaction in which the modes of all three interacting fields
take the form of the Dirac $ \delta $ function. The common
coupling constant $ \tilde{K} $ includes multiplicative factors $
t^\perp f^\parallel $ quantifying the nonlinear interaction in the
medium of length $ L $ and a factor $ \xi_{\rm p}^{({\rm n})} $
allowing to replace the usual pump-field amplitudes (in V/m) by
those expressed in photon numbers; $ \tilde{K} = t^\perp
f^\parallel/(L\xi_{\rm p}^{({\rm n})}) $ (for details, see
\cite{PerinaJr2015a}). The pump-field power $ P $, its repetition
rate $ f $ and its central frequency $ \omega_{\rm p}^0 $
determine the factor $ \xi_{\rm p}^{({\rm n})} $ in the form $
\xi_{\rm p}^{({\rm n})} = \sqrt{ P /(f\hbar\omega_{\rm p}^0)} $.
It is assumed that the pump-field power $ P $ is divided into
individual pump modes $ mlq $ linearly proportionally to their
squared Schmidt coefficients $ (\lambda^\perp_{ml}
\lambda_q^\parallel)^2 $ divided by the squared overlap factors $
\kappa_q^{\parallel 2} $. Such suggested division of the overall
power $ P $ into individual pump modes accords with the
short-length perturbation solution of the Schr\"{o}dinger
equation. This approach assigns an incident classical strong
(coherent) amplitude $ A_{{\rm p},mlq}^{\cal N}(0) \equiv
\tilde{\kappa}^\parallel (\lambda_{ml}^\perp
\lambda_q^\parallel/\kappa_q^\parallel) \xi_{\rm p}^{({\rm n})} $,
$ \tilde{\kappa}^\parallel = 1/ \sqrt{\sum_q \lambda_q^{\parallel
2}/\kappa_q^{\parallel 2}} $, to an $ (mlq)$-th mode.

The momentum operator $ \hat{G}_{\rm int}^{\rm av} $ in
Eq.~(\ref{1}) provides the nonlinear Heisenberg equations written
for individual modes' triplets:
\begin{eqnarray}   
 \frac{ d\hat{a}_{{\rm s},mlq}(z)}{dz} &=&
   K_{mlq}\hat{a}_{{\rm p},mlq}(z) \hat{a}_{{\rm i},mlq}^\dagger(z) , \nonumber \\
 \frac{ d\hat{a}_{{\rm i},mlq}(z)}{dz} &=&
   K_{mlq}\hat{a}_{{\rm p},mlq}(z) \hat{a}_{{\rm s},mlq}^\dagger(z), \nonumber \\
 \frac{ d\hat{a}_{{\rm p},mlq}(z)}{dz} &=&
   -K_{mlq}^* \hat{a}_{{\rm s},mlq}(z) \hat{a}_{{\rm i},mlq}(z).
\label{3}
\end{eqnarray}
In what follows, we pay attention to the dynamics of one typical
modes' triplet and omit the indices $ mlq $ for simplicity. The
Heisenberg equations (\ref{3}) can be solved exactly only for the
interacting fields with small photon numbers using the numerical
approach. For parametric oscillators, the solution of operator
equations linearized around a stationary point can be found
\cite{PerinaJr1993}.

On the other hand, the solution of Eqs.~(\ref{3}) is known
provided that they are considered as classical nonlinear
equations. It can be expressed in terms of elliptic special
functions in general. However, due to the Schmidt decomposition
that symmetrizes the role of the signal and idler fields in the
nonlinear dynamics [$ A_{\rm s}(z) = A_{\rm i}(z) $] we are left
with the following simplified nonlinear equations written for the
real amplitudes $ A_{\rm p} $ and $ A_{\rm s} $ and the real
coupling constant $ K $:
\begin{eqnarray}  
 \frac{dA_{\rm s}(z)}{dz} &=& K A_{\rm p}(z) A_{\rm s}(z) , \nonumber \\
 \frac{dA_{\rm p}(z)}{dz} &=& - K A_{\rm s}^2(z).
\label{4}
\end{eqnarray}
There exists one integral of motion, $ A_{\rm p}^2(z) + A_{\rm
s}^2(z) = A_{\rm p}^2(0) + A_{\rm s}^2(0) \equiv A_{\rm ps}^2 $.
This allows to solve Eqs.~(\ref{4}) as follows
\cite{Bloembergen1965}:
\begin{eqnarray}   
 A_{\rm p}(z) &=& \left[ A_{\rm ps}\frac{A_{\rm p}{\rm cosh}(KA_{\rm ps}z)- A_{\rm ps}{\rm
  sinh}(KA_{\rm ps}z)}{A_{\rm ps}{\rm
  cosh}(KA_{\rm ps}z) - A_{\rm p}{\rm sinh}(KA_{\rm ps}z)} \right],
\label{5}   \\
 A_{\rm s}(z) &=&  \frac{A_{\rm s} A_{\rm ps}}{ A_{\rm ps}{\rm cosh}(KA_{\rm ps}z) -A_{\rm p}{\rm
  sinh}(KA_{\rm ps}z)};
\label{6}
\end{eqnarray}
$ A_{\rm p} \equiv A_{\rm p}(0) $ and $ A_{\rm s} \equiv A_{\rm
s}(0) $. In the case of TWBs that evolve from the signal- and
idler-field vacuum states, we have $ A_{\rm s}(0) = 1/\sqrt{2} $
that corresponds to the symmetric ordering of field operators. We
note that the symmetric ordering is the closest to the classical
solution. In the symmetric ordering, the incident strong pump
field $ A_{\rm p}(0) $ is characterized by the amplitude $ A_{\rm
p}(0) = \sqrt{(A_{\rm p}^{{\cal N}})^2(0) + 1/2} $. The pump field
is gradually depleted by the interaction. At certain point $ z_0 $
the pump field reaches its minimum $ A_{\rm p}(z_0) = A_{\rm s} =
1/\sqrt{2} $ corresponding to the vacuum fluctuations. This occurs
at
\begin{equation}   
 z_0 = \frac{1}{ 2KA_{\rm ps} } \ln \left[ 1 + \frac{2A_{\rm ps}}{A_{\rm ps}+A_{\rm s}}
  \frac{A_{\rm p} - A_{\rm s}}{A_{\rm ps}-A_{\rm p}} \right] .
\label{7}
\end{equation}
At this point, the phases of the interacting fields change such
that the pump field begins to take its energy back from the signal
and idler fields unless $ z = 2z_0 $ where all energy is back in
the pump field. Then the evolution periodically repeats. In the
interval $ z_0 \le z \le 2z_0 $, the fields' evolution is given by
'the mirror symmetry' in which $ z $ is replaced by $ 2z_0 -z $ in
Eqs.~(\ref{5}) and (\ref{6}).

Whereas the classical solution describes well energy (or photon
numbers) and its flow in intense TWBs, it is unable to tackle
their (quantum) statistical properties. For this reason, several
approximate approaches for finding the solution of Eqs.~(\ref{3})
have been suggested and successfully applied. The commonly used
parametric approximation \cite{Gatti2003} considers a strong
un-depleted classical pump field, which linearizes the Heisenberg
equations (\ref{3}). However, it fails in the regime with pump
depletion, in which the so-called generalized parametric
approximation \cite{PerinaJr2016,PerinaJr2016a} allows to find an
appropriate solution. In its framework, the following linear
operator equations are derived:
\begin{eqnarray}  
 \frac{d \hat{a}_{\rm s}(z)}{dz} &=& K A_{\rm p}(z) \hat{a}_{\rm i}^\dagger(z),
  \nonumber \\
 \frac{d \hat{a}_{\rm i}(z)}{dz} &=& K A_{\rm p}(z) \hat{a}_{\rm s}^\dagger(z).
\label{8}
\end{eqnarray}
The classical pump-field amplitude $ A_{\rm p}(z) $ occurring in
Eq.~(\ref{8}) is given in Eq.~(\ref{5}). The solution of
Eqs.~(\ref{8}) is reached in the form:
\begin{eqnarray}  
 \hat{a}_{\rm s}(z) &=& {\rm cosh}[\phi(z)] \hat{a}_{\rm s}(0) +
   {\rm sinh}[\phi(z)] \hat{a}_{\rm i}^\dagger(0) , \nonumber \\
 \hat{a}_{\rm i}(z) &=& {\rm cosh}[\phi(z)] \hat{a}_{\rm i}(0) +
   {\rm sinh}[\varphi(z)] \hat{a}_{\rm s}^\dagger(0),
\label{9}
\end{eqnarray}
where
\begin{equation}   
 \phi(z) = K A_{\rm ps}z - \ln\left[ \frac{A_{\rm ps}+A_{\rm p}}{2A_{\rm ps}} +
  \frac{A_{\rm ps}-A_{\rm p}}{2A_{\rm ps}} \exp(2KA_{\rm ps}z) \right].
\label{10}
\end{equation}
We note that $ \phi(z) = K A_{\rm ps}z $ in the usual parametric
approximation. The generalized parametric approximation, as well
as the usual parametric approximation, does not conserve energy
during the interaction. It also predicts the chaotic statistics of
the emitted strong signal and idler fields. However, as discussed
in the Introduction this is not expected from the physical point
of view as the stimulated emission dominates over the spontaneous
one for the considered strong fields.

For this reason, we consider here a more general momentum operator
$ \hat{G}^{\rm 3p}_{\rm int} $ compared to that written in
Eq.~(\ref{1}):
\begin{eqnarray}     
 \hat{G}_{\rm int}^{\rm 3p}(z) &=& i\hbar
  \sum_{m=-\infty}^{^\infty} \sum_{l,q=0}^{\infty}
   K_{mlq} \left[ A_{{\rm p},mlq}(z) \hat{a}_{{\rm s},mlq}^{\dagger}(z)
   \hat{a}_{{\rm i},mlq}^{\dagger}(z) \right. \nonumber \\
 & & \mbox{} + \gamma A_{{\rm s},mlq}(z) \hat{a}_{{\rm p},mlq}(z)
   \hat{a}_{{\rm i},mlq}^{\dagger}(z) \nonumber \\
 & & \left. \mbox{} + \gamma A_{{\rm i},mlq}(z) \hat{a}_{{\rm p},mlq}(z)
   \hat{a}_{{\rm s},mlq}^{\dagger}(z)  + {\rm h.c.} \right]
\label{11}
\end{eqnarray}
and amplitudes $ A_{{\rm p},mlq}(z) $ and $ A_{{\rm s},mlq}(z)
\equiv A_{{\rm i},mlq}(z) $ are given by the classical solutions
(\ref{5}) and (\ref{6}), respectively. The real constant $ \gamma
$, $ 0\le \gamma \le 1 $, quantifies relative contribution of the
second and third terms in Eq.~(\ref{11}) with respect to the usual
first term (considered in the generalized parametric
approximation). As we will see below, it gives the relative weight
of coherent components in the signal and idler fields of the
emitted TWB. Also, the coupling constants $ K_{mlq} $ are assumed
real.

Concentrating again our attention to an arbitrary mode $ mlq $ the
Heisenberg equations derived from the momentum operator $
\hat{G}^{\rm 3p}_{\rm int} $ attain the form \cite{PerinaJr1993}:
\begin{eqnarray}  
 \frac{d\hat{a}_{\rm s}(z)}{dz} &=& K A_{\rm p}(z) \hat{a}_{\rm i}^\dagger(z) + \gamma K A_{\rm i}(z) \hat{a}_{\rm p}(z) , \nonumber \\
 \frac{d\hat{a}_{\rm i}(z)}{dz} &=& K A_{\rm p}(z) \hat{a}_{\rm s}^\dagger(z) + \gamma K A_{\rm s}(z) \hat{a}_{\rm p}(z) , \nonumber \\
 \frac{d\hat{a}_{\rm p}(z)}{dz} &=& - \gamma K A_{\rm s}(z) \hat{a}_{\rm i}(z) - \gamma K A_{\rm i}(z) \hat{a}_{\rm s}(z) .
\label{12}
\end{eqnarray}
Three equations (\ref{12}) written for the annihilation operators
$ \hat{a}_{b} $, $ b={\rm s,i,p} $, together with their Hermitian
conjugated equations giving the evolution of the creation
operators $ \hat{a}_{b}^\dagger $, $ b={\rm s,i,p} $, form a
closed set of six linear operator equations.

Equations (\ref{12}) can be partially decoupled and transformed
into two two-dimensional sets of equations and two additional
independent equations provided that we introduce the following
quadrature operators:
\begin{equation} 
 \hat{q}_b = \frac{\hat{a}_b + \hat{a}_b^\dagger }{\sqrt{2}},
  \hspace{5mm} \hat{p}_b = \frac{\hat{a}_b - \hat{a}_b^\dagger
  }{i\sqrt{2}}, \hspace{5mm} b={\rm s,i,p},
\label{13}
\end{equation}
and the sum and difference of the signal- and idler-field
operators:
\begin{equation} 
 \hat{q}_{+,-} = \frac{ \hat{q}_{\rm s} \pm \hat{q}_{\rm i} }{\sqrt{2}} ,
 \hspace{5mm}
 \hat{p}_{+,-} =  \frac{ \hat{p}_{\rm s} \pm \hat{p}_{\rm i} }{\sqrt{2}}.
\label{14}
\end{equation}
We note that the introduced operators fulfill the following
commutation relations:
\begin{equation}   
 [\hat{q}_+,\hat{p}_+] = i, \hspace{3mm}
 [\hat{q}_-,\hat{p}_-] = i, \hspace{3mm}
 [\hat{q}_{\rm p},\hat{p}_{\rm p}] = i.
\label{15}
\end{equation}
The commutation relations not written explicitly in Eq.~(\ref{15})
are zero. We reveal the following equations in the newly
introduced operators:
\begin{eqnarray}  
 \frac{d\hat{q}_-(z)}{dz} &=& -K A_{\rm p}(z) \hat{q}_-(z) ,\nonumber \\
 \frac{d\hat{q}_+(z)}{dz} &=& K A_{\rm p}(z) \hat{q}_+(z) + \sqrt{2}\gamma K A_{\rm s}(z) \hat{q}_{\rm p}(z) ,  \nonumber \\
 \frac{d\hat{q}_{\rm p}(z)}{dz} &=& -\sqrt{2} \gamma K A_{\rm s}(z) \hat{q}_+(z) , \nonumber \\
 \frac{d\hat{p}_-(z)}{dz} &=& K A_{\rm p}(z) \hat{p}_-(z) , \nonumber \\
 \frac{d\hat{p}_+(z)}{dz} &=& -K A_{\rm p}(z) \hat{p}_+(z) + \sqrt{2}\gamma K A_{\rm s}(z) \hat{p}_{\rm p}(z) ,  \nonumber \\
 \frac{d\hat{p}_{\rm p}(z)}{dz} &=& - \sqrt{2}\gamma K A_{\rm s}(z) \hat{p}_+(z) .
\label{16}
\end{eqnarray}
Substituting the expressions (\ref{5}) and (\ref{6}) for the
classical amplitudes $ A_{\rm p}(z) $ and $ A_{\rm s}(z) $ in
Eqs.~(\ref{16}) and replacing $ z $ by $ x \equiv KA_{\rm ps}z -
h_{\rm ps} $ with $ h_{\rm ps} \equiv \ln[ (A_{\rm ps}+A_{\rm p})
/ (A_{\rm ps}-A_{\rm p}) ]/2 $ we arrive at the simplified
equations:
\begin{eqnarray}  
 \frac{d\hat{q}_-(x)}{dx} &=& {\rm tanh}(x) \hat{q}_-(x) ,
\label{17}  \\
 \frac{d\hat{q}_+(x)}{dx} &=& - {\rm tanh}(x) \hat{q}_+(x)
  + \frac{\sqrt{2}\gamma }{ {\rm cosh}(x)} \hat{q}_{\rm p}(x)  , \nonumber \\
 \frac{d\hat{q}_{\rm p}(x)}{dx} &=& - \frac{ \sqrt{2} \gamma}{ {\rm cosh}(x)} \hat{q}_+(x) ,
\label{18}  \\
 \frac{d\hat{p}_-(x)}{dx} &=& - {\rm tanh}(x) \hat{p}_-(x) ,
\label{19}  \\
 \frac{d\hat{p}_+(x)}{dx} &=& {\rm tanh}(x) \hat{p}_+(x)
  + \frac{ \sqrt{2}\gamma }{ {\rm cosh}(x)} \hat{p}_{\rm p}(x)  , \nonumber \\
 \frac{d\hat{p}_{\rm p}(x)}{dx} &=& - \frac{ \sqrt{2}\gamma }{ {\rm cosh}(x)}
 \hat{p}_+(x).
\label{20}
\end{eqnarray}

Direct integration of Eqs.~(\ref{17}) and (\ref{19}) leaves us
with the following relations:
\begin{eqnarray}   
 & & \hat{q}_-(x) = {\rm cosh}(x) \hat{q}_-(x=0), \nonumber \\
 & & \hat{p}_-(x) = \frac{1}{{\rm cosh}(x)} \hat{p}_-(x=0).
\label{21}
\end{eqnarray}
The system of Eqs.~(\ref{18}) [(\ref{20})] simplifies when $
\hat{q}_+(x) $ [$ \hat{p}_+(x) $] is expressed as $
\hat{Q}_+(x)/{\rm cosh}(x) $ [$ {\rm cosh}(x) \hat{P}_+(x) $]. We
then arrive at the following equations:
\begin{eqnarray}  
 \frac{d\hat{Q}_+(x)}{dx} &=& \sqrt{2}\gamma  \hat{q}_{\rm p}(x)  , \nonumber \\
 \frac{d\hat{q}_{\rm p}(x)}{dx} &=& - \frac{ \sqrt{2} \gamma}{ {\rm cosh}^2(x)} \hat{Q}_+(x) ,
\label{22}  \\
 \frac{d\hat{P}_+(x)}{dx} &=& \frac{ \sqrt{2}\gamma }{ {\rm cosh}^2(x)} \hat{p}_{\rm p}(x)  , \nonumber \\
 \frac{d\hat{p}_{\rm p}(x)}{dx} &=& - \sqrt{2}\gamma \hat{P}_+(x).
\label{23}
\end{eqnarray}
The solutions of Eqs.~(\ref{22}) and (\ref{23}) can be written for
an arbitrary $ \gamma \in \langle 0,1 \rangle $ in terms of the
hypergeometric functions \cite{Luksxxxx}.

Here, we give only the analytical solution obtained for $ \gamma =
1 $ \cite{Luks1987,Ou1994}:
\begin{eqnarray}   
 & & \left[ \begin{array}{c} \hat{q}_+(x) \\ \hat{q}_{\rm p}(x)
  \end{array} \right] = {\bf F}_{q,x}(x) \left[ \begin{array}{c} \hat{q}_+(x=0)
   \\ \hat{q}_{\rm p}(x=0) \end{array} \right] , \nonumber \\
 & & \left[ \begin{array}{c} \hat{p}_+(x) \\ \hat{p}_{\rm p}(x)
  \end{array} \right] = {\bf F}_{p,x}(x) \left[ \begin{array}{c} \hat{p}_+(x=0)
   \\ \hat{p}_{\rm p}(x=0) \end{array} \right] ,
\label{24}
\end{eqnarray}
where
\begin{eqnarray} 
 {\bf F}_{q,x}^{\gamma=1}(x) &=& \frac{1}{\sqrt{2}{\rm cosh}^2(x)}
  \nonumber \\
 & & \mbox{} \times \left[ \begin{array}{cc} \sqrt{2}[{\rm cosh}(x) -x {\rm sinh}(x)] & 2{\rm
  sinh}(x) \\ -x - {\rm sinh}(x){\rm cosh}(x) & \sqrt{2}
  \end{array} \right] , \nonumber \\
 {\bf F}_{p,x}^{\gamma=1}(x) &=& \frac{1}{\sqrt{2}{\rm cosh}(x)}
   \nonumber \\
 & & \mbox{} \times \left[ \begin{array}{cc} \sqrt{2} & x + {\rm sinh}(x){\rm cosh}(x)  \\
  - 2{\rm sinh}(x) & \sqrt{2}[{\rm cosh}(x) -x {\rm sinh}(x)]
  \end{array} \right] . \nonumber \\
 & &
\label{25}
\end{eqnarray}
We note that for $ \gamma = 0 $ the solution
\begin{eqnarray} 
 {\bf F}_{q,x}^{\gamma=0}(x) &=& \frac{1}{{\rm cosh}(x)}
  \left[ \begin{array}{cc} 1 & 0 \\ 0 &{\rm cosh}(x)
  \end{array} \right] , \nonumber \\
 {\bf F}_{p,x}^{\gamma=0}(x) &=&
  \left[ \begin{array}{cc} {\rm cosh}(x) & 0 \\
  0 & 1  \end{array} \right]  \nonumber \\
 & &
\label{26}
\end{eqnarray}
corresponds to that expressed in Eq.~(\ref{9}).

Returning back to the variable $ z $ giving the position inside
the nonlinear crystal, we arrive at the relations:
\begin{eqnarray}   
 & & \hat{q}_-(z) = f_q(z) \hat{q}_-(z=0), \nonumber \\
 & & \left[ \begin{array}{c} \hat{q}_+(z) \\ \hat{q}_{\rm p}(z)
  \end{array} \right] = {\bf F}_{q}(z) \left[ \begin{array}{c} \hat{q}_+(z=0)
   \\ \hat{q}_{\rm p}(z=0) \end{array} \right] , \nonumber \\
 & & \hat{p}_-(z) = f_p(z) \hat{p}_-(z=0), \nonumber \\
 & & \left[ \begin{array}{c} \hat{p}_+(z) \\ \hat{p}_{\rm p}(z)
  \end{array} \right] = {\bf F}_{p}(z) \left[ \begin{array}{c} \hat{p}_+(z=0)
   \\ \hat{p}_{\rm p}(z=0) \end{array} \right].
\label{27}
\end{eqnarray}
In Eqs.~(\ref{27}), we have
\begin{eqnarray}   
 && f_q(z) = {\rm cosh}(A_{\rm ps}Kz) -\frac{A_{\rm p}}{A_{\rm ps}}
   {\rm sinh}(A_{\rm ps}Kz), \nonumber \\
 && f_p(z) = \frac{1}{f_q(z)} , \nonumber \\
 && {\bf F}_{q}(z) = {\bf F}_{q,x}(KA_{\rm ps}z - h_{\rm ps})
  {\bf F}_{q,x}^{-1}(- h_{\rm ps}) , \nonumber \\
 && {\bf F}_{p}(z) = {\bf F}_{p,x}(KA_{\rm ps}z - h_{\rm ps})
  {\bf F}_{p,x}^{-1}(- h_{\rm ps}) .
\label{28}
\end{eqnarray}

Inverse transformations to those given in Eqs.~(\ref{13}) and
(\ref{14}) applied in Eqs.~(\ref{27}) allow us to arrive at the
expressions giving the evolution of annihilation and creation
operators:
\begin{equation}  
 \hat{\bf a}(z) = {\bf U}(z) \hat{\bf a}(z=0) + {\bf V}(z) \hat{\bf
  a}^\dagger(z=0).
\label{29}
\end{equation}
In Eq.~(\ref{29}), the vectors $ \hat{\bf a}^T \equiv [
\hat{a}_{\rm s}, \hat{a}_{\rm i}, \hat{a}_{\rm p} ] $ and $
\hat{\bf a}^{\dagger T} \equiv [ \hat{a}_{\rm s}^\dagger,
\hat{a}_{\rm i}^\dagger, \hat{a}_{\rm p}^\dagger ] $ are
conveniently introduced and symbol $ T $ denotes transposition.
The matrices $ {\bf U} $ and $ {\bf V} $ defined as
\begin{eqnarray} 
 {\bf U}(z) &=& \frac{1}{2} \left[ {\bf M}_q(z) + {\bf M}_p(z)
 \right] , \nonumber \\
 {\bf V}(z) &=& \frac{1}{2} \left[ {\bf M}_q(z) - {\bf M}_p(z)
 \right]
\label{30}
\end{eqnarray}
are expressed in terms of the matrices $ {\bf M}_q $ and $ {\bf
M}_p $:
\begin{eqnarray} 
 && {\bf M}_b(z) = \frac{1}{2} \left[ \begin{array}{ccc}
  f_b + F_{b,11} & -f_b + F_{b,11} & \sqrt{2} F_{b,12} \\
  -f_b + F_{b,11} & f_b + F_{b,11} & \sqrt{2} F_{b,12} \\
  \sqrt{2} F_{b,21} & \sqrt{2} F_{b,21} & 2 F_{b,22} \end{array}
  \right], \nonumber \\
 && \hspace{3cm} b=q,p.
\label{31}
\end{eqnarray}
We note that the solution (\ref{29}) preserves the canonical
commutation relations. As a consequence the matrices $ {\bf U} $
and $ {\bf V} $ obey the following relations:
\begin{eqnarray}  
 {\bf U}(z) {\bf V}^T(z) - {\bf V}(z) {\bf U}^T(z) &=& {\bf 0} ,
  \nonumber \\
 {\bf U}(z) {\bf U}^T(z) - {\bf V}(z) {\bf V}^T(z) &=& {\bf 1} ;
\label{32}
\end{eqnarray}
$ {\bf 0} $ ($ {\bf 1} $) stands for the zero (unity)
3-dimensional matrix.

\section{Statistical properties of the modes in individual triplets}

Statistical properties of the interacting signal, idler and pump
modes in a triplet are conveniently described by the normal
characteristic function $ C_{\cal N} $ \cite{Perina1991}
determined for the incident statistical operator $ \varrho(z=0) $
along the relation
\begin{eqnarray}   
 C_{\cal N}(\mbox{\boldmath$\beta$},z) &=& {\rm Tr} \left\{ \exp[
 \mbox{\boldmath$\beta$}^T {\bf a}^\dagger(z) ] \exp[-\mbox{\boldmath$\beta$}^{*T} {\bf
  a}(z)] \varrho(z=0) \right\} ; \nonumber \\
 & &
\label{33}
\end{eqnarray}
$ \mbox{\boldmath$\beta$}^T \equiv (\beta_{\rm s},\beta_{\rm
i},\beta_{\rm p}) $. Assuming the signal and idler modes in the
incident vacuum states and the pump mode in an incident coherent
state corresponding to the classical field, we can express the
normal characteristic function $ C_{\cal N} $ as follows
\cite{Perina1991}:
\begin{eqnarray}   
 C_{\cal N}(\mbox{\boldmath$\beta$},z) &=& \exp\left[ \sum_{j={\rm s,i,p}}
  \left[ -B_j(z)|\beta_j|^2 + {\rm Re}\left\{ C_j(z)\beta_j^{*2} \right\}
  \right] \right] \nonumber \\
 & & \hspace{-12mm} \times \exp\left[ \sum_{jk={\rm si, sp, ip}}
  2{\rm Re} \left\{ D_{jk}(z)\beta_j^*\beta_k^* + \bar{D}_{jk}(z)\beta_j\beta_k^* \right\} \right] \nonumber \\
 & & \times \exp\left[ 2i {\rm Im} \left\{\mbox{\boldmath$\xi$}^{*T}(z)\mbox{\boldmath$\beta$} \right\} \right] ;
\label{34}
\end{eqnarray}
$ {\rm Re} $ ($ {\rm Im} $) denotes the real (imaginary) part of
the argument. In Eq.~(\ref{34}), the coherent amplitudes $
\mbox{\boldmath$\xi$} \equiv (\xi_{\rm s},\xi_{\rm i},\xi_{\rm p})
$ are given as
\begin{equation}  
 \mbox{\boldmath$\xi$}(z) = {\bf U}(z)\mbox{\boldmath$\xi$}(z=0) +
  {\bf V}(z)\mbox{\boldmath$\xi$}^*(z=0),
\label{35}
\end{equation}
and functions $ B_j $, $ C_j $, $ D_{jk} $, and $ \bar{D}_{jk} $
are derived in terms of the real evolution matrices $ {\bf U} $
and $ {\bf V } $:
\begin{eqnarray} 
 B_j(z) &=& \sum_{k={\rm s,i,p}} V_{jk}^2(z) , \nonumber \\
 C_j(z) &=& \sum_{k={\rm s,i,p}} U_{jk}(z) V_{jk}(z) , \nonumber \\
 D_{jk}(z) &=& \sum_{l={\rm s,i,p}} U_{jl}(z)V_{kl}(z) , \nonumber \\
 \bar{D}_{jk}(z) &=& -\sum_{k={\rm s,i,p}} V_{jl}(z)V_{kl}(z) .
\label{36}
\end{eqnarray}

Using the normal characteristic function $ C_{\cal N} $ written in
Eq.~(\ref{34}), the mean intensity $ i_j \equiv \langle
\hat{i}_j\rangle  = \langle \hat{a}_j^\dagger \hat{a}_j \rangle $
of mode $ j $, $ j={\rm s,i,p} $, is obtained in the form:
\begin{equation} 
 i_j(z) = B_j(z) + |\xi_j(z)|^2,
\label{37}
\end{equation}
where the first (second) term on the r.h.s. of Eq.~(\ref{37})
characterizes the chaotic (coherent) part of the field. Similarly,
the mean squared intensity fluctuation $ \langle (\Delta
\hat{i}_j)^2 \rangle \equiv \langle \hat{a}_j^{\dagger 2}
\hat{a}_j^2 \rangle - \langle \hat{a}_j^\dagger \hat{a}_j
\rangle^2 $ of mode $ j $ is derived as follows:
\begin{equation} 
 \langle[\Delta \hat{i}_j(z)]^2 \rangle = b_j^2(z) + | c_j(z)|^2
  - 2\left|\xi_j(z)\right|^4 ;
\label{38}
\end{equation}
$ b_j(z) \equiv B_j(z) + |\xi_j(z)|^2 $, $ c_j(z) \equiv C_j(z) +
\xi_j^2(z)$. Correlation of intensity fluctuations $ \langle
\Delta \hat{i}_j \Delta \hat{i}_k \rangle \equiv \langle
\hat{a}_j^\dagger\hat{a}_k^\dagger \hat{a}_j \hat{a}_k\rangle -
\langle \hat{a}_j^\dagger \hat{a}_j \rangle \langle
\hat{a}_k^\dagger \hat{a}_k \rangle $ of modes $ j $ and $ k $ is
determined in the form:
\begin{equation} 
 \langle\Delta \hat{i}_j(z) \Delta \hat{i}_k(z) \rangle = | d_{jk}(z)|^2 + | \bar{d}_{jk}(z)|^2
  - 2\left|\xi_j(z)\right|^2 \left|\xi_k(z)\right|^2;
\label{39}
\end{equation}
$ d_{jk}(z) \equiv D_{jk}(z) + \xi_j(z)\xi_k(z) $ and $
\bar{d}_{jk}(z) \equiv -\bar{D}_{jk}(z) + \xi_j^*(z)\xi_k(z) $.

To quantify the type of statistics and correlations of the fields
it is useful to define the dimensionless reduced intensity moments
$ r_j $ and intensity-fluctuation moments $ r_{jk} $ along the
relations:
\begin{eqnarray}  
 r_j(z) &=& \frac{ \langle\hat{i}_j^2(z) \rangle }{
 \langle\hat{i}_j(z) \rangle^2},
\label{40}   \\
 r_{jk}(z) &=& \frac{ \langle\Delta\hat{i}_j(z) \Delta\hat{i}_k(z) \rangle }{
  \langle\hat{i}_j(z) \rangle \langle\hat{i}_k(z) \rangle} .
\label{41}
\end{eqnarray}
The signal-idler sub-shot-noise intensity correlations are
described by parameter $ R_{\rm si} $ defined in terms of the
moments of intensities as (for details, see \cite{Perina2007}):
\begin{equation}  
 R_{\rm si}(z) = 1 + \frac{ \langle [\hat{i}_{\rm s}(z) - \hat{i}_{\rm i}(z)]^2 \rangle }{
 \langle\hat{i}_{\rm s}(z) \rangle + \langle\hat{i}_{\rm i}(z)\rangle } .
\label{42}
\end{equation}
Nonclassical TWBs are characterized by the values of parameter $
R_{\rm si} $ lower than 1, i.e. $ \langle (\hat{i}_{\rm s} -
\hat{i}_{\rm i})^2 \rangle < 0 $ for such TWBs. Phase squeezing in
mode $ j $ is judged according to the value of principal squeeze
variance $ \lambda $ \cite{Luks1988} given as
\begin{equation}  
 \lambda_j(z) = 1/2 + B_j(z) - \left|C_j(z)\right| .
\label{43}
\end{equation}
For phase squeezed states, the principal squeeze variance $
\lambda $ attains values lower than 1/2.

\section{Statistical properties of the interacting fields and
their correlations}

The signal, idler and pump fields are composed of many independent
modes belonging to individual modes' triplets. According to the
theory presented in \cite{PerinaJr2015,PerinaJr2015a}, the modes
are considered as a direct product of the spectral modes (indexed
by $ q $) and the modes defined in the wave-vector transverse
plane (indexed by $ ml $). Before we apply the results of the
previous section, we have to generalize them to include a random
phase $ \varphi $ originating in the incident signal and idler
vacuum states \cite{Graham1968,Graham1968a,PerinaJr1993}. The
classical solution in modes' triplet $ mlq $ does not necessarily
have to be real. The signal and idler electric-field amplitudes
can attain randomly a phase $ \varphi_{mlq} $ such that $ A_{{\rm
s},mlq} \rightarrow A_{{\rm s},mlq}\exp(i\varphi_{mlq}) $ and  $
A_{{\rm i},mlq} \rightarrow A_{{\rm i},mlq}\exp(-i\varphi_{mlq})
$; i.e. the signal- and idler-field phases compensate each other.
The same applies for the operator solution in a given triplet
where we have $ \hat a_{{\rm s},mlq} \rightarrow \hat a_{{\rm
s},mlq} \exp(i\varphi_{mlq}) $ and $ \hat a_{{\rm i},mlq}
\rightarrow \hat a_{{\rm i},mlq} \exp(-i\varphi_{mlq}) $. As a
consequence, the amplitudes $ \xi_b $, $ b={\rm s,i} $, in
Eq.~(\ref{35}) and functions $ B_b $, $ C_b $, $ D_{\rm si} $ and
$ \bar{D}_{\rm si} $ in Eq.~(\ref{36}) are modified as follows
\begin{eqnarray}  
 \xi_{{\rm s},mlq}^\varphi &=& \xi_{{\rm s},mlq} \exp(i\varphi_{mlq}), \nonumber \\
 \xi_{{\rm i},mlq}^\varphi &=& \xi_{{\rm i},mlq} \exp(-i\varphi_{mlq}), \nonumber \\
 B_{{\rm s},mlq}^\varphi &=& B_{{\rm s},mlq}, \nonumber \\
 B_{{\rm i},mlq}^\varphi &=& B_{{\rm i},mlq}, \nonumber \\
 C_{{\rm s},mlq}^\varphi &=& C_{{\rm s},mlq}\exp(2i\varphi_{mlq}), \nonumber \\
 C_{{\rm i},mlq}^\varphi &=& C_{{\rm i},mlq}\exp(-2i\varphi_{mlq}), \nonumber \\
 D_{{\rm si},mlq}^\varphi &=& D_{{\rm si},mlq}, \nonumber \\
 \bar{D}_{{\rm si},mlq}^\varphi &=& \bar{D}_{{\rm si},mlq}\exp(-2i\varphi_{mlq}).
\label{44}
\end{eqnarray}
The ensemble averages for physical quantities characterizing the
multi-mode fields are then obtained after averaging over the
phases $ \varphi_{mlq} $ with the uniform distributions $
p(\varphi_{mlq}) = 1/(2\pi) $. These averages are indicated by
subscript $ \varphi $.

The mean intensity $ I_j $ of field $ j $, $ j={\rm s,i,p} $, is
obtained as the sum of its coherent $ \langle \hat{I}_j^{\rm c}
\rangle_\varphi $ and chaotic $ \langle \hat{I}_j^{\rm ch}
\rangle_\varphi $ components:
\begin{eqnarray}   
 & I_j \equiv \langle \hat{I}_j \rangle_\varphi = \langle \hat{I}_j^{\rm c} \rangle_\varphi
  + \langle \hat{I}_j^{\rm ch} \rangle_\varphi , &
\label{45} \\
 & \langle \hat{I}_j^{\rm c} \rangle_\varphi = \sum_{mlq}
 |\xi_{j,mlq}|^2 , \hspace{5mm} \langle \hat{I}_j^{\rm ch} \rangle_\varphi = \sum_{mlq}
  B_{j,mlq}. & \nonumber
\end{eqnarray}
The mean quadratic intensity fluctuation $ \langle
(\Delta\hat{I}_j)^2 \rangle_\varphi $ of field $ j $ is derived
along the formula:
\begin{equation}  
 \langle(\Delta\hat{I}_j)^2 \rangle_\varphi = \sum_{mlq} \Bigl[
  b^2_{j,mlq} + |c_{j,mlq}|^2 - 2|\xi_{j,mlq}|^4 \Bigr] .
\label{46}
\end{equation}
The mean cross-correlation of intensity fluctuations $ \langle
\Delta\hat{I}_j \Delta\hat{I}_k \rangle_\varphi $ of fields $ j $
and $ k $ is expressed as:
\begin{equation}  
 \langle \Delta\hat{I}_j \Delta\hat{I}_k \rangle_\varphi = \sum_{mlq} \Bigl[
  |d_{jk,mlq}|^2 + |\bar{d}_{j,mlq}|^2 - 2|\xi_{j,mlq}|^4 \Bigr] .
\label{47}
\end{equation}

In analogy to the definitions in Eqs.~(\ref{40}---\ref{42}), we
define parameter $ \tilde{R}_{\rm si} $ to quantify the
signal-idler sub-shot-noise intensity correlations and reduced
intensity moments $ \tilde{r}_j $ and intensity-fluctuation
moments $ \tilde{r}_{jk} $ for fields $ j $ and $ k $:
\begin{eqnarray}  
 \tilde{R}_{\rm si} &=& 1 + \frac{\langle (\hat{I}_{\rm s} - \hat{I}_{\rm i} )^2 \rangle_\varphi}{
 \langle\hat{I}_{\rm s} \rangle_\varphi + \langle\hat{I}_{\rm i}\rangle_\varphi } ,
\label{48}   \\
 \tilde{r}_j &=& \frac{ \langle\hat{I}_j^2 \rangle_\varphi }{ \langle\hat{I}_j \rangle^2_\varphi},
\label{49}   \\
 \tilde{r}_{jk} &=& \frac{ \langle\Delta\hat{I}_j \Delta\hat{I}_k \rangle_\varphi }{ \langle\hat{I}_j
  \rangle_\varphi \langle\hat{I}_k\rangle_\varphi} .
\label{50}
\end{eqnarray}
The intensity moments also allow for the determination of an
effective number $ K^{\rm n} $ of modes constituting the TWB
\cite{Perina1985}:
\begin{equation}  
 K^{\rm n} = \frac{ \langle\hat{I}_j \rangle_\varphi^2 }{\langle(\Delta\hat{I}_j)^2
 \rangle_\varphi}.
\label{51}
\end{equation}
Alternatively, we may quantify the number $ K $ of modes via the
'mean photon-pair amplitudes' $ \langle \hat{a}_{{\rm s},mlq}
\hat{a}_{{\rm i},mlq} \rangle_\varphi $ as follows
\cite{PerinaJr2013}:
\begin{eqnarray}  
 & & K = \frac{ \left( \sum_{mlq} |\langle \hat{a}_{{\rm s},mlq} \hat{a}_{{\rm i},mlq}
 \rangle_\varphi|^2 \right)^2 }{ \sum_{mlq} |\langle \hat{a}_{{\rm s},mlq} \hat{a}_{{\rm i},mlq}
 \rangle_\varphi|^4 } ,
\label{52} \\
 & & \langle \hat{a}_{{\rm s},mlq} \hat{a}_{{\rm i},mlq} \rangle_\varphi
  = d_{{\rm si},mlq} .
\nonumber
\end{eqnarray}

The spectral intensity (cross-) correlations are described by the
following intensity-fluctuation  auto- ($ A_{j,\omega} $) and
cross- ($ C_{\omega} $) correlation functions:
\begin{eqnarray}  
 A_{j,\omega}(\omega_j,\omega'_j) &=& \langle :
  \Delta[\hat{a}_j^\dagger(\omega_j) \hat{a}_j(\omega_j) ]
  \Delta[\hat{a}_j^\dagger(\omega'_j) \hat{a}_j(\omega'_j)
  ]: \rangle_{\perp,\varphi} , \nonumber \\
 & & \hspace{30mm} \hspace{5mm} j={\rm s,i},
\label{53} \\
 C_{\omega}(\omega_{\rm s},\omega_{\rm i}) &=& \langle :
  \Delta[\hat{a}_{\rm s}^\dagger(\omega_{\rm s}) \hat{a}_{\rm s}(\omega_{\rm s}) ]
  \Delta[\hat{a}_{\rm i}^\dagger(\omega_{\rm i}) \hat{a}_{\rm i}(\omega_{\rm i})
  ]: \rangle_{\perp,\varphi}, \nonumber \\
 & &
\label{54}
\end{eqnarray}
where symbol $ :: $ denotes the normal ordering of field operators
and subscript $ \perp $ means averaging over the transverse modes.
The intensity-fluctuation correlation functions $ A_{j,\omega} $
and $ C_{\omega} $ are expressed in terms of the Schmidt spectral
modes $ f_{j,q} $, $ j={\rm s,i} $, in the form (for details, see
\cite{PerinaJr2015a}):
\begin{eqnarray}  
 A_{j,\omega}(\omega_j,\omega'_j) &=& \sum_{ml} \left| \sum_q
  f_{j,q}^*(\omega_j) f_{j,q}(\omega'_j)
  b_{j,mlq} \right|^2 \nonumber \\
 & & \mbox{} + \sum_{mlq} \left| f_{j,q}(\omega_j) f_{j,q}(\omega'_j)
   c_{j,mlq}\right|^2 \nonumber \\
 & & \mbox{} - 2\sum_{mlq} \left| f_{j,q}(\omega_j) f_{j,q}(\omega'_j)\right|
  |\xi_{j,mlq}|^4,
\label{55} \\
 C_{\omega}(\omega_{\rm s},\omega_{\rm i}) &=& \sum_{ml} \left| \sum_q
  f_{{\rm s},q}(\omega_{\rm s}) f_{{\rm i},q}(\omega_{\rm i})
  d_{{\rm si},mlq} \right|^2 \nonumber \\
 & & \mbox{} \hspace{-8mm} + \sum_{mlq} \left| f_{{\rm s},q}^*(\omega_{\rm s}) f_{{\rm i},q}(\omega_{\rm i})
   \bar{d}_{{\rm si},mlq}\right|^2 \nonumber \\
 & & \mbox{} \hspace{-8mm} - 2\sum_{mlq} \left| f_{{\rm s},q}(\omega_{\rm s}) f_{{\rm i},q}(\omega_{\rm i})\right|
  |\xi_{{\rm s},mlq}\xi_{{\rm i},mlq}|^2 .
\label{56}
\end{eqnarray}

Also the pump-field spectral intensity $ I_{\rm p}^{\rm tr} $
transferred into the down-converted fields can approximately be
determined along the formula relying on the 'mean photon-pair
amplitudes':
\begin{eqnarray}   
 I_{\rm p}^{\rm tr}(\omega_{\rm p}) &=& \sum_{ml} \left|
  \int_{-\infty}^{\infty}d\omega_{\rm s}
  \int_{-\infty}^{\infty}d\omega_{\rm i} \,
  \delta(\omega_{\rm p} - \omega_{\rm s}-\omega_{\rm i}) \right. \nonumber \\
 & & \mbox{} \times \left. \langle
   \hat{a}_{{\rm s},ml}(\omega_{\rm s}) \hat{a}_{{\rm i},ml}(\omega_{\rm i})
   \rangle_{\varphi} \right|^2.
\label{57}
\end{eqnarray}
In Eq.~(\ref{57}), the annihilation operators are written in
frequencies $ \omega $ (for details, see \cite{PerinaJr2015a}).
Using the pump-field modes $ f_{{\rm p},q}^{({\rm n})} $ defined
in Eq.~(\ref{2}), the intensity $ I_{\rm p}^{\rm tr} $ is
expressed as follows:
\begin{equation} 
 I_{\rm p}^{\rm tr}(\omega_{\rm p}) = \sum_{ml} \left| \sum_q
   f_{{\rm p},q}^{({\rm n})}(\omega_{\rm p}) d_{{\rm si},mlq}\right|^2.
\label{58}
\end{equation}

\section{Sum-frequency generation and Hong-Ou-Mandel interference}

Correlations between the signal and idler fields manifest
themselves also in the time domain, where they allow for the
observation of the coherent components. To describe temporal
properties of TWBs, we first write the spatial and spectral
positive-frequency operator amplitudes $
\hat{E}_{b,\omega}^{(+)}({\bf k}_b^\perp,\omega_b) $ belonging to
a monochromatic plane wave of field $ b $ with frequency $
\omega_b $ and transverse wave vector $ {\bf k}_b^\perp $, $
b={\rm s,i} $:
\begin{equation}   
 \hat{E}_{b,\omega}^{(+)}({\bf k}_b^\perp,\omega_b) = i
 \sqrt{\frac{\hbar\omega_b}{2\varepsilon_0 c}} \sum_{mlq}
  t_{b,ml}({\bf k}_b^\perp) f_{b,q}(\omega_b) \hat{a}_{b,mlq}.
\label{59}
\end{equation}
In Eq.~(\ref{59}), the Schmidt-mode functions $ t_{b,ml} $
describe the fields in their transverse planes similarly as the
already used Schmidt-mode functions $ f_{b,q} $ are applied in the
frequency domain (for details, see \cite{PerinaJr2015a}). The
positive-frequency operator amplitudes $ \hat{E}_b^{(+)}({\bf
k}_b^\perp,t_b) $ appropriate in the time domain are then
expressed in terms of the temporal Schmidt modes $ \tilde{f}_{b,q}
$ as:
\begin{eqnarray}   
 \hat{E}_b^{(+)}({\bf k}_b^\perp,t_b) &=& \frac{i}{\sqrt{2\varepsilon_0 c}} \sum_{mlq}
  t_{b,ml}({\bf k}_b^\perp) \tilde{f}_{b,q}(t_b) \hat{a}_{b,mlq} ;
\label{60}  \\
 & & \hspace{-20mm} \tilde{f}_{b,q}(t) = \frac{1}{\sqrt{2\pi}} \int
  d\omega_b \sqrt{\frac{\omega_b}{\omega_b^0}} f_{b,q}(\omega_b) \exp(-i\omega_b
  t);
\label{61}
\end{eqnarray}
$ \omega_b^0 $ stands for the central frequency of field $ b $.

Photon flux $ I_{b,t} $ of field $ b $ expressed in photon numbers
is then determined as follows:
\begin{eqnarray}  
 I_{b,t} &=& \frac{2\varepsilon_0 c}{\hbar\omega_b^0} \langle \hat{E}_b^{(-)}(t)\hat{E}_b^{(+)}(t)
  \rangle_{\perp,\varphi} \nonumber \\
 &=& \sum_{ml} \sum_{q} | \tilde{f}_{b,q}(t)|^2 b_{b,mlq}, \hspace{5mm} b={\rm s,i} .
\label{62}
\end{eqnarray}
Temporal intensity auto- ($ A_{b,t} $) and cross-correlation ($
C_{t} $) functions are given by the formulas analogous to those in
Eqs.~(\ref{55}) and (\ref{56}) derived for the spectral
correlation functions.

The process of sum-frequency generation \cite{Boyd2003} represents
the basic tool in the experimental analysis of temporal
correlations between the signal and idler fields
\cite{Jedrkiewicz2012}. In the method, the signal and idler fields
are mutually delayed by time delay $ \tau $ and then they generate
a sum-frequency field in a nonlinear crystal with $ \chi^{(2)} $
nonlinearity. Assuming perfect phase matching in the crystal,
intensity $ I^{\rm SFG} $ of the sum-frequency field averaged over
the transverse plane is given as:
\begin{eqnarray}   
 I^{\rm SFG}(\tau) &=& \nonumber \\
 & & \hspace{-17mm} \eta
  \int_{-\infty}^{\infty} dt \langle \hat{E}_{\rm s}^{(-)}(t+\tau)
  \hat{E}_{\rm i}^{(-)}(t) \hat{E}_{\rm s}^{(+)}(t+\tau) \hat{E}_{\rm i}^{(+)}(t)
  \rangle_{\perp,\varphi} \hspace{5mm}
\label{63}
\end{eqnarray}
where $ \eta $ is a suitable constant linearly proportional to the
squared $ \chi^{(2)} $ susceptibility. Applying formulas
(\ref{60}) and (\ref{61}), relation (\ref{63}) can be rearranged
into the form:
\begin{eqnarray}  
 I^{\rm SFG}(\tau) &=& \frac{\eta \hbar^2\omega_{\rm s}^0\omega_{\rm i}^0}{4\varepsilon_0^2c^2}
 \int_{-\infty}^{\infty} dt \sum_{ml} w_{ml}
  \Biggl\{ \nonumber \\
 & & \sum_q|\tilde{f}_{{\rm s},q}(t+\tau)|^2 b_{{\rm s},mlq}
   \sum_{q'} |\tilde{f}_{{\rm i},q'}(t)|^2 b_{{\rm i},mlq'}  \nonumber \\
 & & \mbox{} + \left|\sum_q \tilde{f}_{{\rm s},q}(t+\tau) \tilde{f}_{{\rm i},q}(t) d_{{\rm si},mlq} \right|^2  \nonumber \\
 & & \mbox{} + \sum_q \left| \tilde{f}^*_{{\rm s},q}(t+\tau) \tilde{f}_{{\rm i},q}(t) \bar{d}_{{\rm si},mlq}\right|^2
  \nonumber \\
 & & \mbox{} - 2\sum_q |\tilde{f}_{{\rm s},q}(t+\tau) \tilde{f}_{{\rm i},q}(t)|^2 |\xi_{{\rm s},mlq} \xi_{{\rm i},mlq}|^2
  \Biggr\} .  \nonumber \\
\label{64}
\end{eqnarray}
The weights $ w_{ml} $ of transverse modes characterize the
nonlinear spatial overlap of the signal and idler modes. They are
defined as
\begin{equation} 
 w_{ml}  = \int_{0}^{2\pi} d\varphi \int_{0}^{\infty} dr r
  |t_{{\rm s},ml}(r,\varphi)t_{{\rm i},ml}(r,\varphi)|^2
\label{65}
\end{equation}
using the signal and idler transverse mode functions $ t_{\rm s} $
and $ t_{\rm i} $, respectively, written in the radial
coordinates. We assume that the TWB in the crystal output plane is
imaged into the area of a thin nonlinear crystal used for the
sum-frequency generation.

Temporal correlations in TWBs can also be experimentally analyzed
in the Hong-Ou-Mandel interferometer \cite{Hong1987} that is based
upon mixing the mutually delayed signal and idler fields (by time
delay $ \tau $) at a balanced beam splitter and simultaneous
detection of intensities at both output ports of the beam
splitter. The number $ R $ of 'coincident' intensity detections is
given by the formula
\begin{eqnarray}   
 R(\tau) &=& \frac{4\varepsilon_0^2 c^2}{\hbar^2\omega_{\rm s}^0\omega_{\rm i}^0} \int_{-\infty}^{\infty} dt_{\rm A} \int_{-\infty}^{\infty} dt_{\rm B} \nonumber \\
 & & \langle \hat{E}_{\rm A}^{(-)}(t_{\rm A}) \hat{E}_{\rm B}^{(-)}(t_{\rm B}) \hat{E}_{\rm A}^{(+)}(t_{\rm A}) \hat{E}_{\rm B}^{(+)}(t_{\rm B})
  \rangle_{\perp,\varphi} , \hspace{5mm}
\label{66}
\end{eqnarray}
where the operator amplitudes $ \hat{E}_{\rm A}^{(+)} $ and $
\hat{E}_{\rm B}^{(+)} $ at the output ports $ {\rm A} $ and $ {\rm
B} $, respectively, are given as:
\begin{eqnarray}   
 \hat{E}_{\rm A}^{(+)}(t) &=& r \hat{E}_{\rm i}^{(+)}(t)
  + t \hat{E}_{\rm s}^{(+)}(t-\tau) , \nonumber \\
 \hat{E}_{\rm B}^{(+)}(t) &=& t^* \hat{E}_{\rm i}^{(+)}(t)
  - r^* \hat{E}_{\rm s}^{(+)}(t-\tau) .
\label{67}
\end{eqnarray}
Symbol $ r $ ($ t $) in Eq.~(\ref{67}) stands for the
beam-splitter reflectivity (transmissivity). Formula (\ref{66})
for the number $ R $ of 'coincident' intensity detections can be
rewritten to include averaging over the transverse plane:
\begin{eqnarray}   
 R(\tau) &=& \frac{4\varepsilon_0^2 c^2}{\hbar^2\omega_{\rm s}^0\omega_{\rm i}^0}
  \int_{-\infty}^{\infty} dt_{\rm A} \int_{-\infty}^{\infty} dt_{\rm B} \nonumber \\
 & & \hspace{-11mm} \Bigl\{ \langle :\Delta\left[ \hat{E}_{\rm A}^{(-)}(t_{\rm A})\hat{E}_{\rm A}^{(+)}(t_{\rm A})\right]
  \Delta\left[\hat{E}_{\rm B}^{(-)}(t_{\rm B}) \hat{E}_{\rm B}^{(+)}(t_{\rm B}) \right]:
  \rangle_{\perp,\varphi} \nonumber \\
 & & \hspace{-11mm} \mbox{} +
 \langle\hat{E}_{\rm A}^{(-)}(t_{\rm A})\hat{E}_{\rm A}^{(+)}(t_{\rm A})\rangle_{\perp,\varphi}
  \langle\hat{E}_{\rm B}^{(-)}(t_{\rm B})\hat{E}_{\rm B}^{(+)}(t_{\rm B})\rangle_{\perp,\varphi}
  \Bigr\} .
\label{68}
\end{eqnarray}
Moreover, it is useful to determine the interference pattern
formed by the intensity fluctuations $ \Delta I $ instead of only
intensities $ I $:
\begin{eqnarray}   
 R^\Delta(\tau) &=& \frac{4\varepsilon_0^2 c^2}{\hbar^2\omega_{\rm s}^0\omega_{\rm i}^0}
  \int_{-\infty}^{\infty} dt_{\rm A} \int_{-\infty}^{\infty} dt_{\rm B} \nonumber \\
 & & \hspace{-11mm} \langle :\Delta\left[ \hat{E}_{\rm A}^{(-)}(t_{\rm A})\hat{E}_{\rm A}^{(+)}(t_{\rm A})\right]
  \Delta\left[\hat{E}_{\rm B}^{(-)}(t_{\rm B}) \hat{E}_{\rm B}^{(+)}(t_{\rm B}) \right]:
  \rangle_{\perp,\varphi} . \nonumber \\
 & &
\label{69}
\end{eqnarray}

The normalized intensity and intensity-fluctuation correlation
functions $ R_{\rm n} $ and $ R_{\rm n}^\Delta $ derived from
Eqs.~(\ref{68}) and (\ref{69}), respectively, are expressed as
follows:
\begin{eqnarray}  
 R_{\rm n}(\tau) &=& 1 - \frac{ 2{\rm Re} \{ \varrho^\Delta(\tau) \} }{
  R_0 +R_0^\Delta } ,
\label{70} \\
 R_{\rm n}^\Delta(\tau) &=& 1 - \frac{ 2{\rm Re} \{\varrho^\Delta(\tau) \} }{ R_0^\Delta } .
\label{71}
\end{eqnarray}
In Eqs.~(\ref{70}) and (\ref{71}), the complex interference term $
\varrho^\Delta(\tau) $ is given as
\begin{eqnarray}  
 \varrho^\Delta(\tau) &=& \sum_{ml} \Bigl\{ |rt|^2 \Bigl[
  -2 \sum_{q} |g_{qq}(\tau)|^2 |\xi_{{\rm s},mlq}\xi_{{\rm i},mlq}|^2
  \nonumber \\
 & & \mbox{} + \sum_{qq'} g_{qq'}(\tau) g^*_{q'q}(\tau) d^*_{{\rm si},mlq}
  d_{{\rm si},mlq'} \nonumber \\
 & & \mbox{} + \sum_{qq'} |g_{qq'}(\tau)|^2 b_{{\rm i},mlq}b_{{\rm s},mlq'} \Bigr]
  \Bigr\}
\label{72}
\end{eqnarray}
using the signal-idler mode amplitude correlation functions $
g_{qq'} $ defined as
\begin{equation}  
 g_{qq'}(\tau) = \int_{-\infty}^{\infty} d\omega f^*_{{\rm i},q}(\omega)
 f_{{\rm s},q'}(\omega) \exp(i\omega\tau).
\label{73}
\end{equation}
The normalization constants $ R_0 $ and $ R_0^\Delta $ introduced
in Eqs.~(\ref{70}) and (\ref{71}) are obtained in the form:
\begin{eqnarray}  
 R_0 &=& |rt|^2 \Bigl[ \Bigl(\sum_{mlq} b_{{\rm i},mlq}\Bigr)^2 +
  \Bigl(\sum_{mlq} b_{{\rm s},mlq}\Bigr)^2 \Bigr] \nonumber \\
 & & \mbox{} + (|r|^4 +|t|^4)
  \sum_{mlq} b_{{\rm i},mlq} \sum_{m'l'q'} b_{{\rm s},m'l'q'} ,
\label{74} \\
 R_0^\Delta &=& \sum_{ml} \Bigl\{ |rt|^2 \sum_{q} (-2|\xi_{{\rm i},mlq}|^4 + b_{{\rm i},mlq}^2 +
  |c_{{\rm i},mlq}|^2 \nonumber \\
 & & \mbox{} -2|\xi_{{\rm s},mlq}|^4 + b_{{\rm s},mlq}^2 + |c_{{\rm s},mlq}|^2 ) + (|r|^4 +|t|^4)
  \nonumber \\
 & & \mbox{} \times \sum_{q} (-2|\xi_{{\rm s},mlq}\xi_{{\rm i},mlq}|^2 + |d_{{\rm si},mlq}|^2 +
  |\bar{d}_{{\rm si},mlq}|^2 ) \Bigr\} . \nonumber \\
 & &
\label{75}
\end{eqnarray}

\section{Behavior of individual modes' triplets}

To demonstrate the behavior of the analyzed model at the
elementary level, we first analyze the properties of individual
modes grouped into a typical triplet. We consider a 4-mm long BBO
crystal cut for non-collinear type-I process for the
spectrally-degenerate interaction pumped by a pulse at the
wavelength $ \lambda_{\rm p}^0 = 349 $~nm with spectral width $
\Delta\lambda_{\rm p} = 1 $~nm (FWHM), transverse profile with
radius $ w_{\rm p} = 500 $~$ \mu $m and repetition rate $ f = 400
$~s$ {}^{-1} $ impinging on the crystal at normal incidence (for
more details, see \cite{PerinaJr2015a}). As an example, we
investigate the modes' properties of the triplet with the greatest
initial pump power. This mode is characterized by the greatest
Schmidt coefficient $ \lambda_{\rm max} $ and overlap factor $
\kappa_{\rm max}^\parallel $ that give $ \tilde{K}\kappa_{\rm
max}^\parallel L = 23.29 $~W$^{-1/2} $ for the crystal of length $
L = 4 $~mm. When the pump-power axis $ P $ is rescaled to $
|\lambda_{\rm max}\kappa^\parallel/(\lambda\kappa_{\rm
max}^\parallel )|^2 P $ the results valid for an arbitrary triplet
with the Schmidt coefficient $ \lambda $ and overlap factor $
\kappa^\parallel $ are obtained.

The inclusion of the second and third terms in the momentum
operator $ \hat{G}^{\rm 3p}_{\rm int} $ in Eq.~(\ref{11}) results
in the generation of coherent components in the signal and idler
fields, that are initially in the vacuum state. This is
accompanied by the occurrence of a chaotic component in the pump
field. The comparison of curves in Fig.~1 giving the intensities
of the signal ($ i_{\rm s} $) and pump ($ i_{\rm p} $) fields
shows that they depend only weakly on the parameter $ \gamma $
quantifying the relative weight of the second and third terms in
the momentum operator $ \hat{G}^{\rm 3p}_{\rm int} $.
\begin{figure}         
 \resizebox{0.8\hsize}{!}{\includegraphics{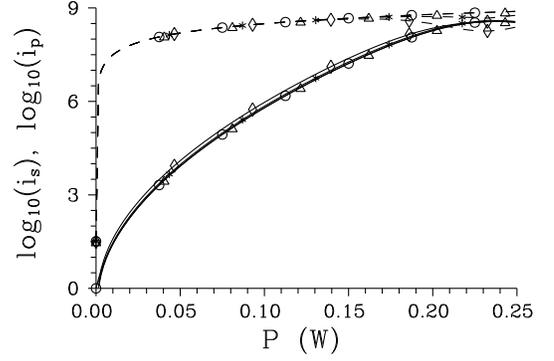}}
 \caption{Signal ($ i_{\rm s} $, solid curves) and pump ($ i_{\rm p} $, dashed curves)
  intensities (in photon numbers) as they depend on pump power $ P
  $ for $ \gamma = 1 $ ($ \diamond $), $ \gamma = 0.5 $ ($ \ast
  $), $ \gamma = 0.1 $ ($ \triangle $), and $ \gamma = 0 $ ($ \circ
  $).}
\end{figure}
The curves plotted in Fig.~1 show that the increase of
signal-field intensity $ i_{\rm s} $ with the increasing pump
power $ P $ stops at around $ P_{\rm th} \approx 240 $~mW where
the nonlinear process switches its nonlinear phase and, as a
consequence, decrease of the signal-field intensity $ i_{\rm s} $
follows. The relative weight $ c_{\rm s} $ of the signal-field
intensity $ i^{\rm c}_{\rm s} $ of the coherent component with
respect to the overall intensity $ i_{\rm s} $ is decisive for the
properties of TWBs or, more precisely, for the declination of
their properties from those characterizing the usual chaotic TWBs
($ \gamma = 0 $). The curves drawn in Fig.~2 show that the weight
$ c_{\rm s} $ of the signal mode monotonically increases with the
increasing parameter $ \gamma $ and is larger than 30~\% for $
\gamma = 1 $.
\begin{figure}         
 \resizebox{0.8\hsize}{!}{\includegraphics{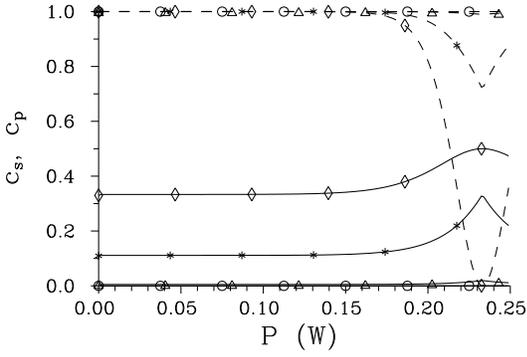}}
 \caption{Relative weights $ c_{\rm s} $ (solid curves) and $ c_{\rm p} $ (dashed curves)
  of intensities of the coherent components in the signal and pump
  modes, respectively, as they depend on pump power $ P
  $ for $ \gamma = 1 $ ($ \diamond $), $ \gamma = 0.5 $ ($ \ast
  $), $ \gamma = 0.1 $ ($ \triangle $), and $ \gamma = 0 $ ($ \circ
  $).}
\end{figure}
Also, the weight $ c_{\rm s} $ naturally increases with the pump
power $ P $. On the other hand, the weight $ c_{\rm p} $ of the
pump mode decreases when both parameter $ \gamma $ and pump power
$ P $ increase. For $ \gamma = 1 $, the incident coherent pump
mode is even transformed into a purely chaotic mode for the
threshold pump power $ P_{\rm th} $. Whereas the presence of the
coherent component in the signal mode causes the mode's
declination from its incident chaotic statistics ($ r = 2 $)
towards the coherent ones ($ r=1 $), the chaotic component of the
pump mode is too weak to significantly change the mode's character
(see Fig.~3).
\begin{figure}         
 \resizebox{0.8\hsize}{!}{\includegraphics{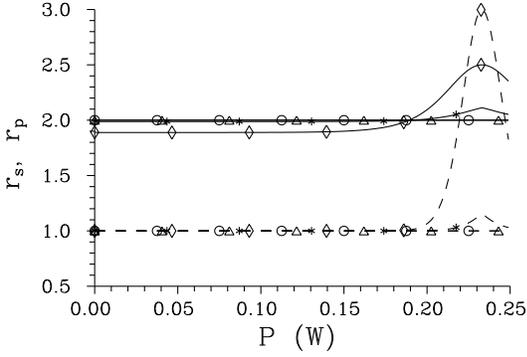}}
 \caption{Reduced intensity moments of the signal ($ r_{\rm s} $, solid curves) and pump
  ($ c_{\rm p} $, dashed curves) modes as they depend on pump power $ P
  $ for $ \gamma = 1 $ ($ \diamond $), $ \gamma = 0.5 $ ($ \ast
  $), $ \gamma = 0.1 $ ($ \triangle $), and $ \gamma = 0 $ ($ \circ
  $).}
\end{figure}
This behavior qualitatively changes for pump powers $ P $ around $
P_{\rm th} $ for which all three modes attain their super-chaotic
statistics ($ r>2 $), due to the intense energy exchange. This is
a purely quantum effect. It is accompanied by the generation of
phase-squeezed light \cite{Dodonov2002} in the pump mode as the
curves giving the principal squeeze variance $ \lambda_{\rm p} $
in Fig.~4 confirm.
\begin{figure}         
 \resizebox{0.8\hsize}{!}{\includegraphics{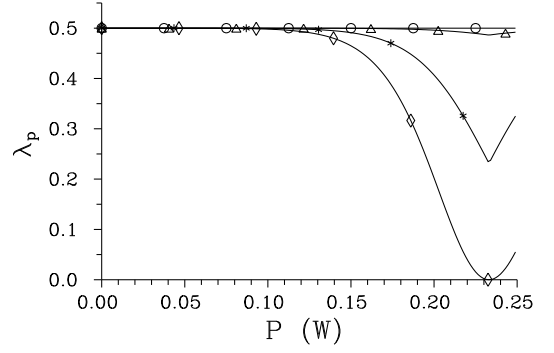}}
 \caption{Principal squeeze variance $ \lambda_{\rm p} $ of the pump mode as it depends on pump power $ P
  $ for $ \gamma = 1 $ ($ \diamond $), $ \gamma = 0.5 $ ($ \ast
  $), $ \gamma = 0.1 $ ($ \triangle $), and $ \gamma = 0 $ ($ \circ
  $).}
\end{figure}

The three-mode interaction naturally generates correlations among
the intensities of the interacting modes. When $ \gamma = 0 $, the
signal and idler photons are emitted solely in pairs which results
in ideal sub-shot-noise correlations between the signal- and
idler-field intensities. The second and third terms in the
momentum operator $ \hat{G}^{\rm 3p}_{\rm int} $ partially break
photon pairing which leads to the loss of ideal sub-shot-noise
intensity correlations. However, the sub-shot-noise intensity
correlations still exist at lower pump powers $ P $, as documented
in Fig.~5.
\begin{figure}         
 \resizebox{0.8\hsize}{!}{\includegraphics{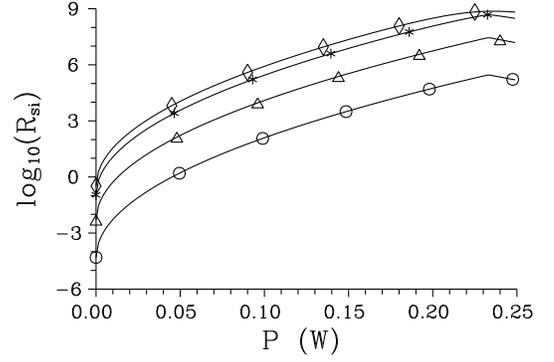}}
 \caption{Parameter $ R_{\rm si} $ quantifying the signal-idler sub-shot-noise
  intensity correlations as it depends on pump power $ P
  $ for $ \gamma = 1 $ ($ \diamond $), $ \gamma = 0.5 $ ($ \ast
  $), $ \gamma = 0.1 $ ($ \triangle $), and $ \gamma = 0.01 $ ($ \circ
  $).}
\end{figure}
It holds that the smaller the parameter $ \gamma $ the wider the
area of pump powers $ P $ in which the sub-shot-noise intensity
correlations are observed. When no sub-shot-noise intensity
correlations are found, the classical intensity correlations
quantified by the reduced signal-idler intensity-fluctuation
moment $ r_{\rm si} $ still exist (see Fig.~6).
\begin{figure}         
 \resizebox{0.8\hsize}{!}{\includegraphics{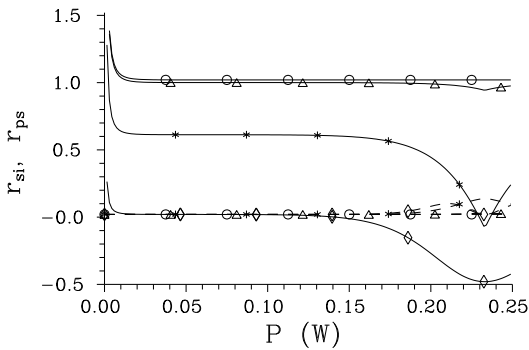}}
 \caption{Reduced intensity-fluctuation moments $ r_{\rm si} $ (solid curves) and $ r_{\rm ps} $ (dashed curves)
  quantifying the signal-idler and pump-signal intensity correlations, respectively, as they depend on pump power $ P
  $ for $ \gamma = 1 $ ($ \diamond $), $ \gamma = 0.5 $ ($ \ast
  $), $ \gamma = 0.1 $ ($ \triangle $), and $ \gamma = 0 $ ($ \circ
  $).}
\end{figure}
If the parameter $ \gamma \approx 1 $, the signal-idler
intensity-fluctuation correlations are observed only for weak pump
powers $ P $. On the other hand, the signal-idler
intensity-fluctuation anti-correlations
\cite{Mista1969,Perinova1981} are developed for pump powers $ P $
around $ P_{\rm th} $. The transition from intensity-fluctuation
correlations to anti-correlations as the pump power $ P $
increases can be understood from the form of momentum operator $
\hat{G}^{\rm 3p}_{\rm int} $ written in Eq.~(\ref{11}) as follows.
The first term in Eq.~(\ref{11}) dominates for lower pump powers $
P $. It generates photons in pairs, i.e. with ideal intensity
(photon number) correlations between the signal and idler modes.
As it is well known, phases of the signal- and idler-field
amplitudes differ only in their signs in this case
\cite{Graham1968}. On the other hand, the second and third terms
in Eq.~(\ref{11}) are decisive for the properties of TWBs at
greater pump powers $ P $. These terms couple the signal and idler
modes via the pump mode and they enforce synchronization of the
signal- and idler-field phases. As a consequence, the intensity
fluctuations of the signal and idler modes have to be mutually
anti-correlated to comply with the 'phase-intensity uncertainty
relations'. The dominance of the second and third terms in the
momentum operator $ \hat{G}^{\rm 3p}_{\rm int} $ for the powers $
P $ around $ P_{\rm th} $ also provides correlations between the
signal- (idler-) and pump-field intensity fluctuations (see
Fig.~6).

\section{Behavior of the interacting fields}

We discuss properties of the whole TWB emphasizing the role of
coherent fields' components. The analyzed TWB is composed of
several hundred thousand spatio-spectral signal and idler Schmidt
modes (for the behavior of chaotic TWBs, see
\cite{PerinaJr2015a,PerinaJr2016}) that evolve due to the energy
provided by the pump field. Each pair of the signal and idler
modes is accompanied by its own pump mode. Due to the relation
among the modes in an arbitrary modes' triplet baaed on the
convolution, the pump modes exhibit faster oscillations in their
intensity spectral profiles compared to their signal and idler
counterparts. In case of the interaction symmetric with respect to
the exchange of the signal and idler fields, which is considered
here, the number of local minima in the pump intensity profile is
twice the number of those in the signal (or idler) intensity
profile. The signal and pump profiles of the first three spectral
modes are compared in Fig.~7.
\begin{figure}         
 \resizebox{.45\hsize}{!}{\includegraphics{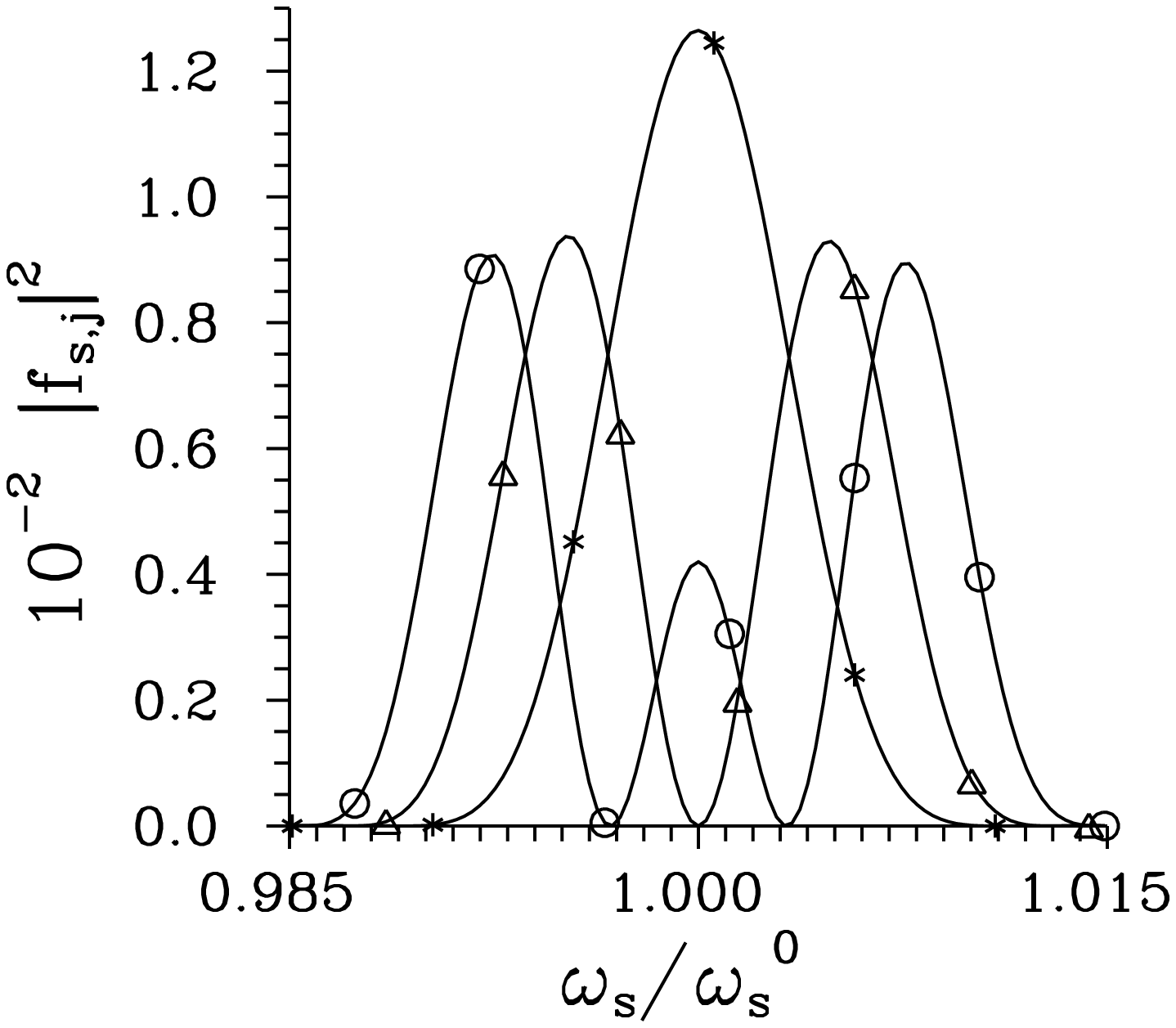}}
  \hspace{2mm}
 \resizebox{.45\hsize}{!}{\includegraphics{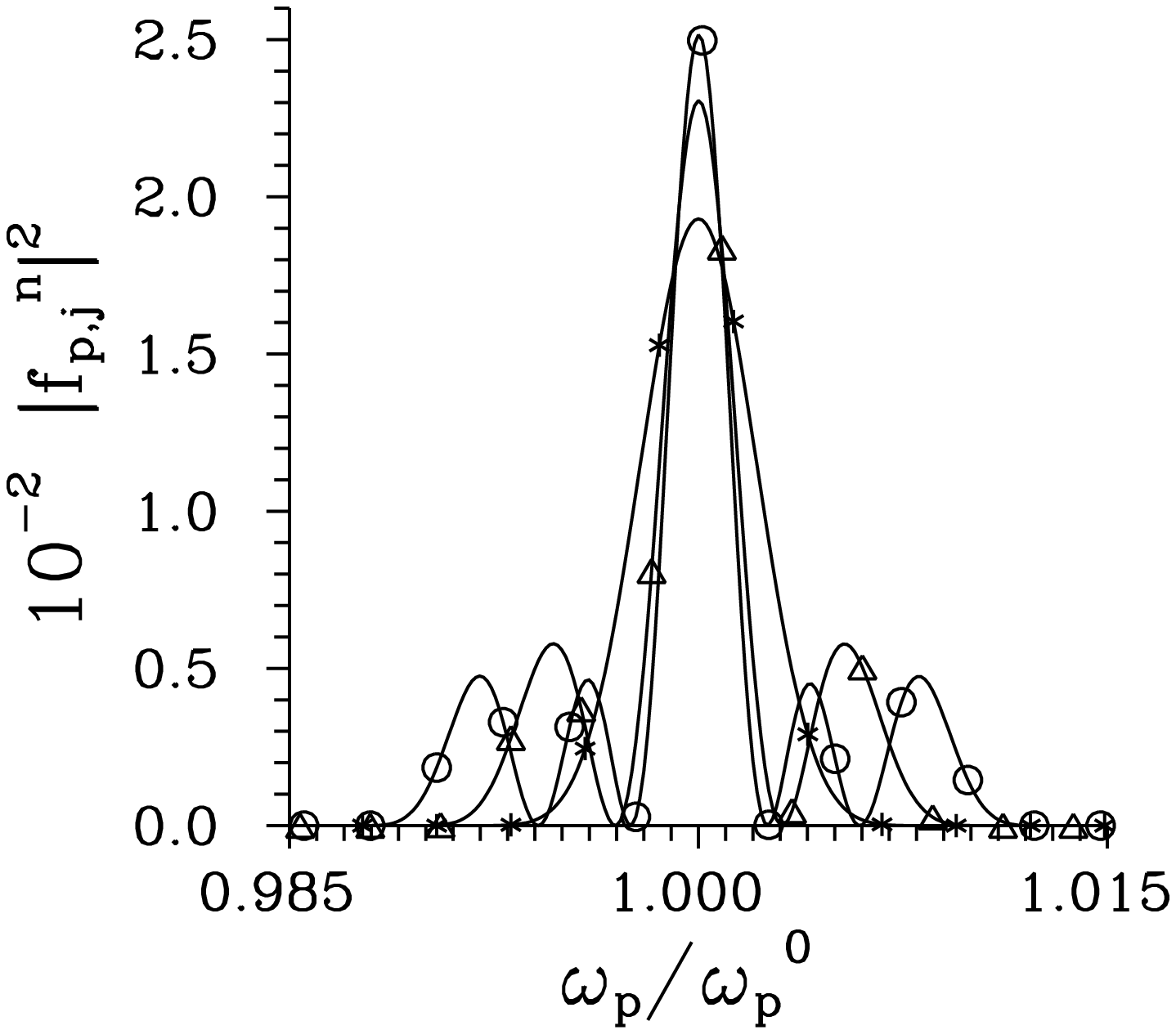}}

 \centerline{ (a) \hspace{.4\hsize} (b) }
 \caption{(a) Signal ($ |f_{{\rm s},j}|^2 $) and
  (b) pump ($ |f_{{\rm p},j}^{({\rm n})}|^2 $) spectral intensity profiles for the first three modes;
  $ j=1 $ ($ \ast $), 2 ($ \triangle $) and 3 ($ \circ $). The profiles are normalized such that
  $ \int d\omega_b|f_{b,j}(\omega_b)|^2 /\omega_b^0 = 1 $, $ b={\rm s,p} $.}
\end{figure}
In the interaction, depletion of the pump modes may occur inside
the crystal and even the back-flow of energy from the signal and
idler modes into the corresponding pump mode is observed for
sufficiently strong pump fields. This results in the complex
dynamics of multi-mode TWBs, as we discuss below.

Owing to pump depletion, the exponential growth of the
signal-field intensity $ I_{\rm s} $ with the increasing pump
power $ P $ is replaced by a roughly linear growth for greater
pump powers $ P $, independently on the presence of the coherent
fields' components [see Fig.~8(a)].
\begin{figure}         
 \resizebox{.45\hsize}{!}{\includegraphics{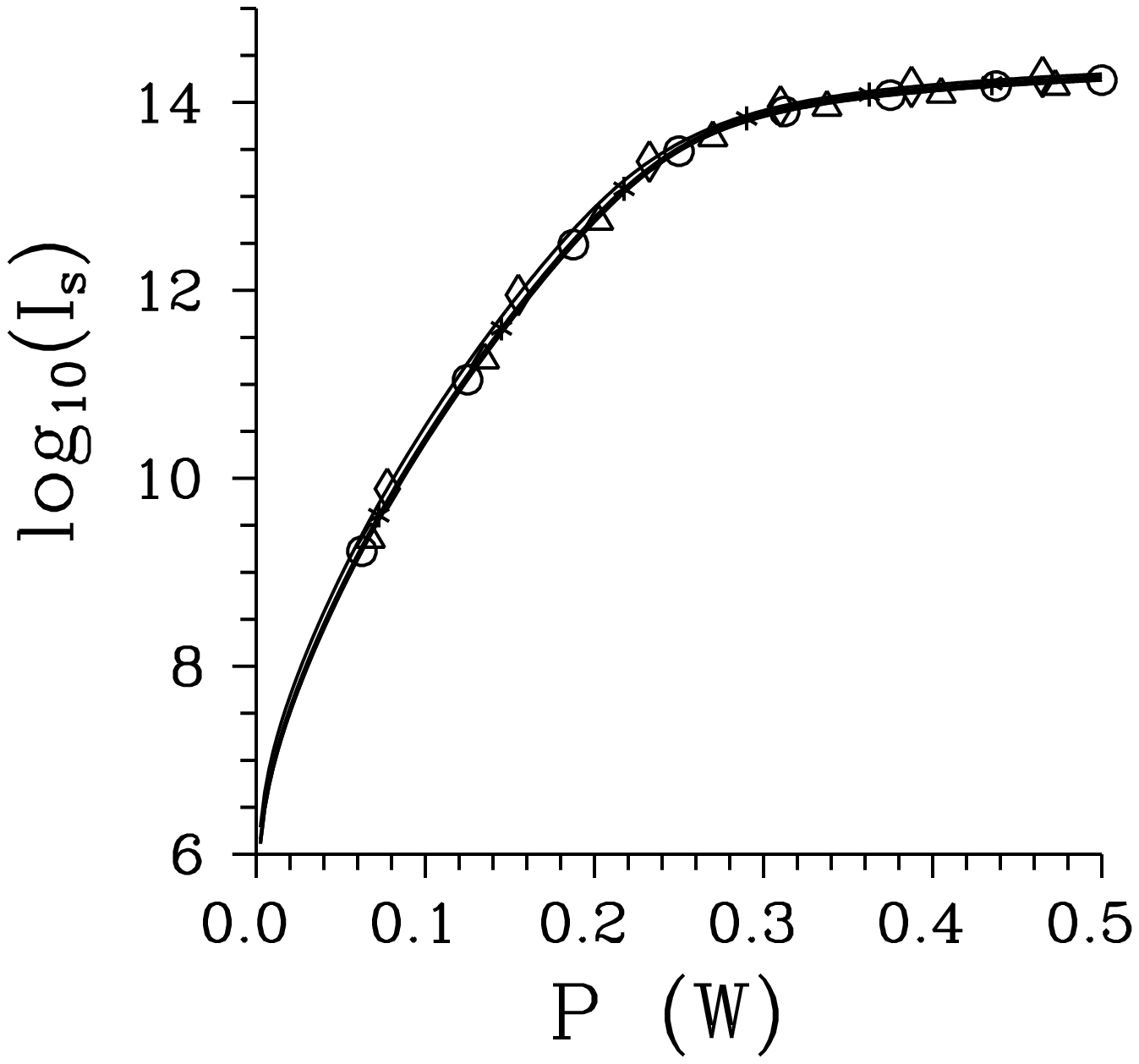}}
  \hspace{2mm}
 \resizebox{.45\hsize}{!}{\includegraphics{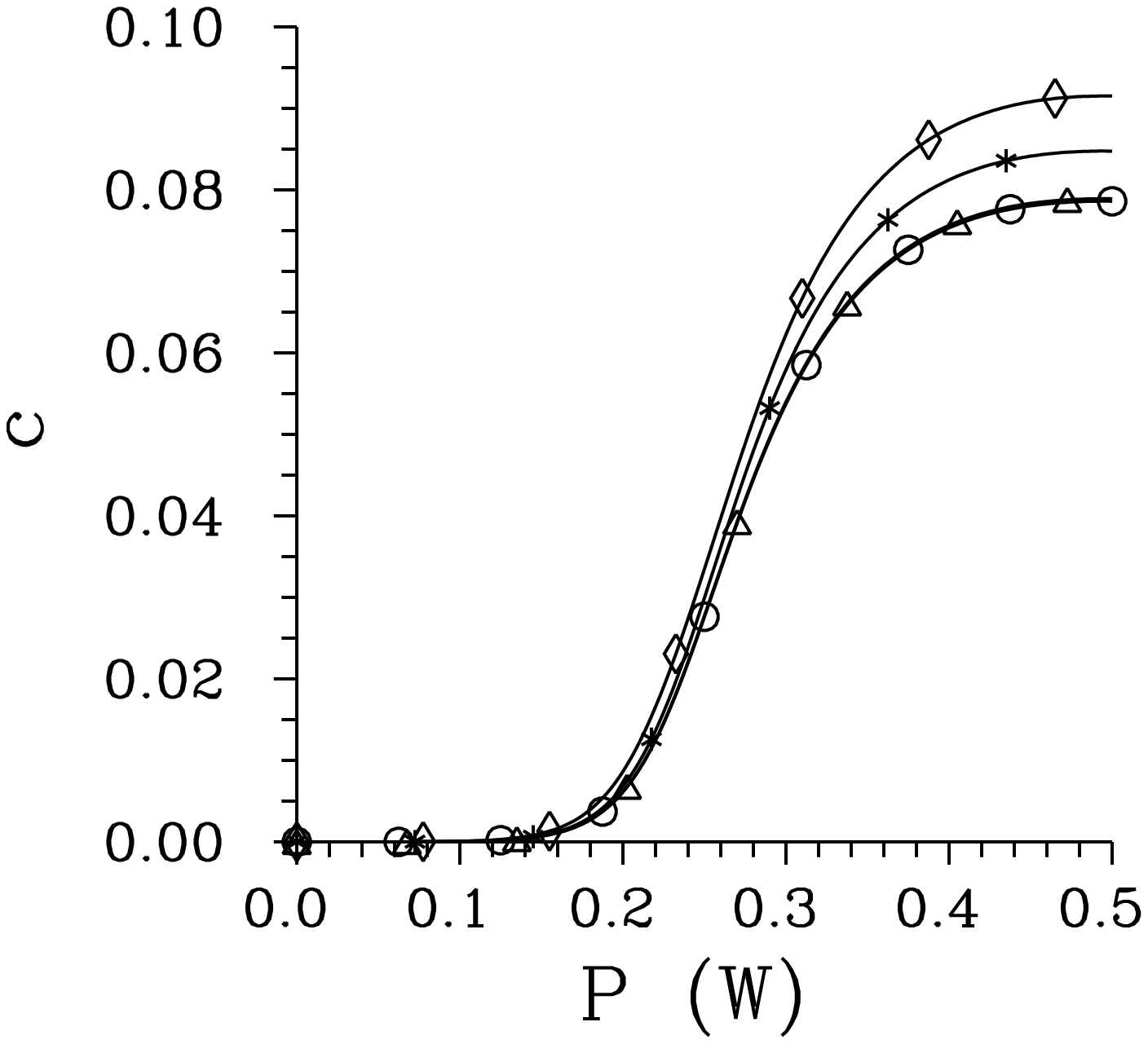}}

 \centerline{ (a) \hspace{.4\hsize} (b) }
 \caption{(a) Mean signal intensity $ I_{\rm s} $ (in photon numbers) and
  (b) conversion efficiency $ c $ giving the relative energy converted from the pump field [$ c \equiv
  I_{\rm s}(z=L)/I_{\rm p}(z=0) $] as they depend on pump power $ P
  $ for $ \gamma = 1 $ ($ \diamond $), $ \gamma = 0.5 $ ($ \ast
  $), $ \gamma = 0.1 $ ($ \triangle $), and $ \gamma = 0 $ ($ \circ $).}
\end{figure}
On the other hand, the coherent fields' components allow for the
faster exchange of energy between the pump and the down-converted
modes, which results in larger conversion efficiencies $ c $ drawn
in Fig.~8(b). As follows from the curves plotted in Fig.~8(b), the
conversion efficiencies $ c $ approach 9~\% for pump powers $ P $
greater than 400~mW. The inclusion of the second and third terms
in the momentum operator $ \hat{G}^{\rm 3p}_{\rm int} $ in
Eq.~(\ref{11}) results in intense coherent components in the
signal and idler fields. The coherent components play an important
role for all pump powers $ P $. The relative contribution $ c_{\rm
s}^{\rm c} $ of the signal coherent component exceeds 40~\% for
greater pump powers $ P $ and $ \gamma =1 $, as documented in
Fig.~9(a).
\begin{figure}         
 \resizebox{.45\hsize}{!}{\includegraphics{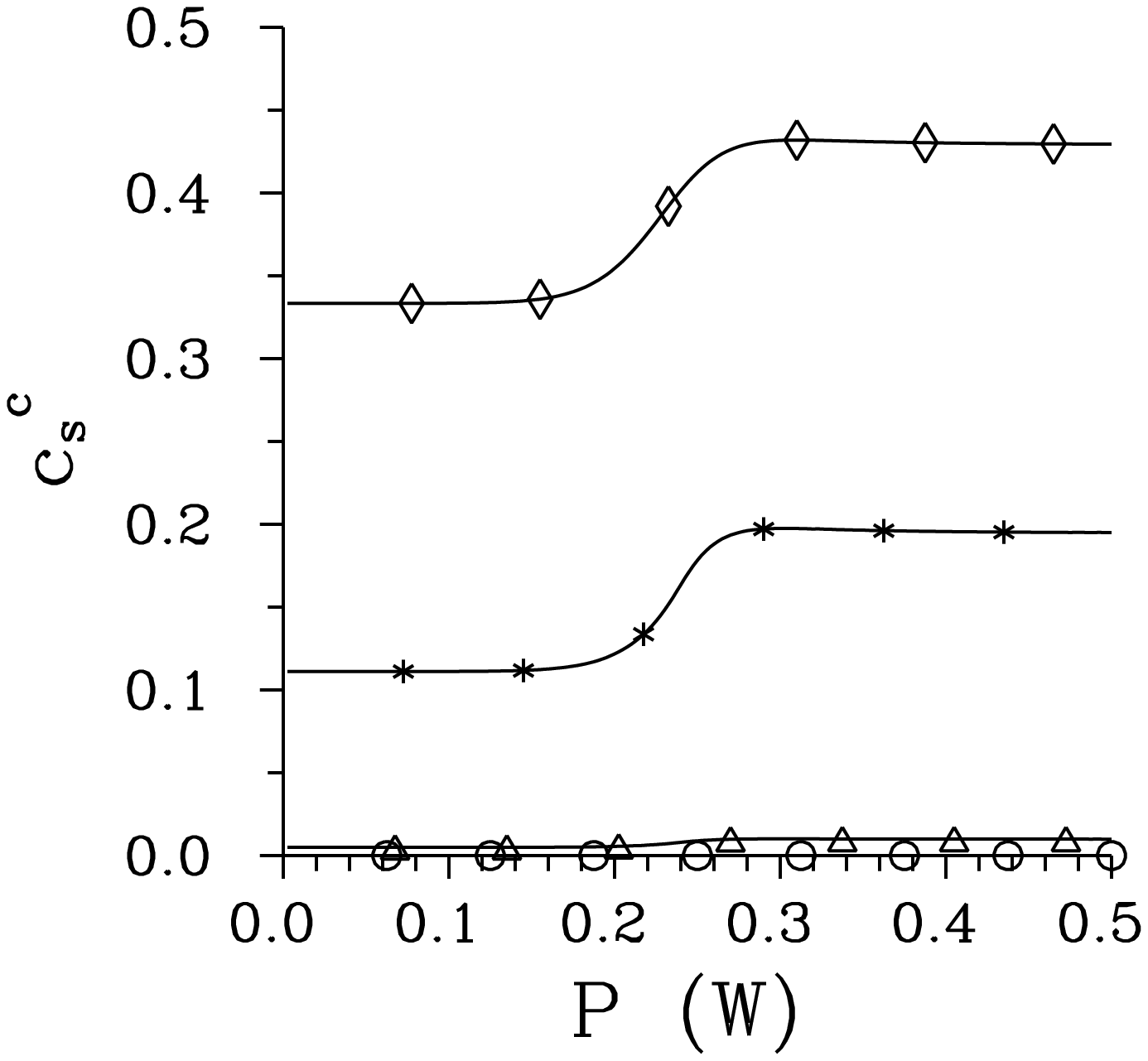}}
  \hspace{2mm}
 \resizebox{.45\hsize}{!}{\includegraphics{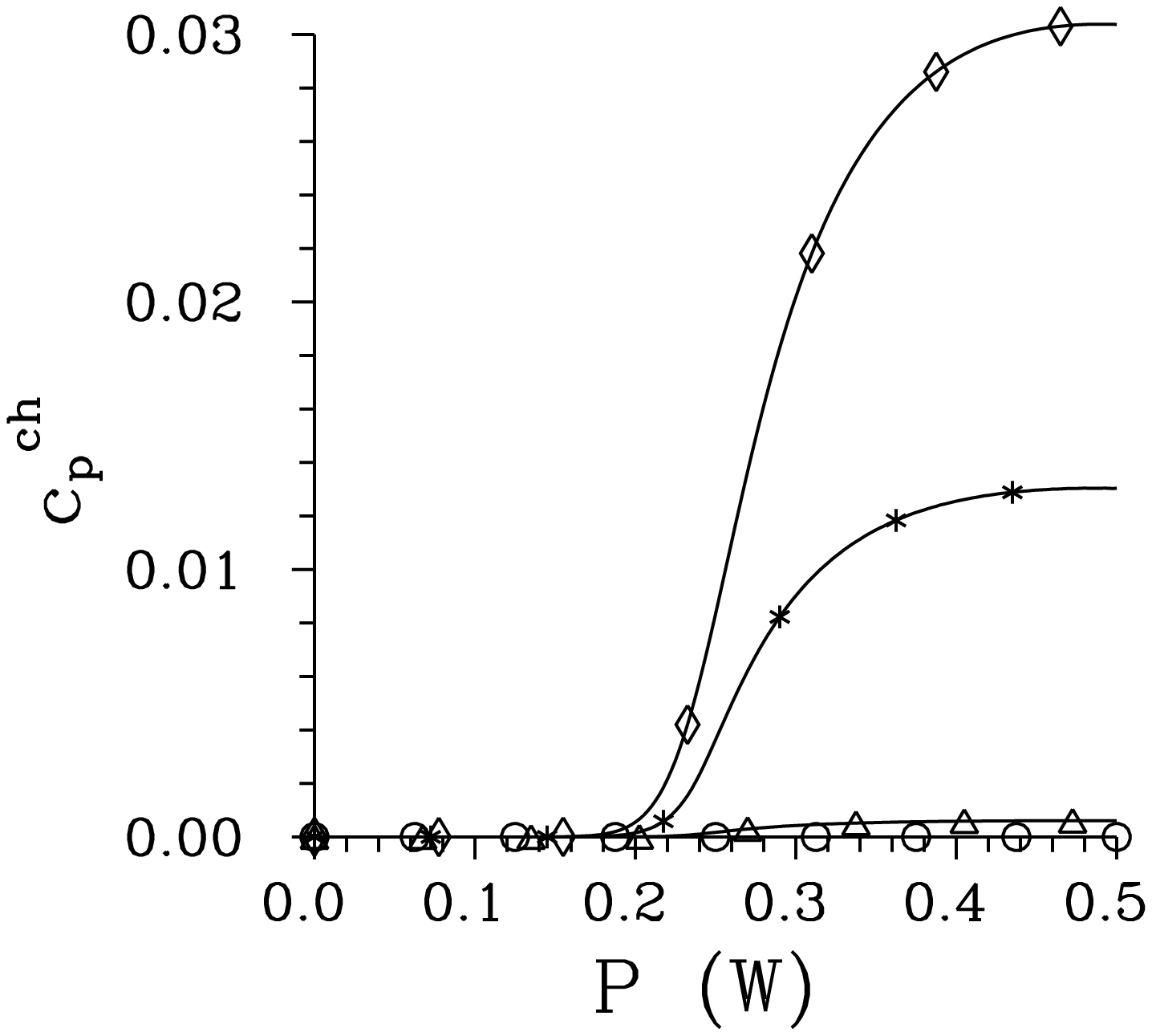}}

 \centerline{ (a) \hspace{.4\hsize} (b) }
 \caption{Relative contribution of (a) the signal coherent component $ c_{\rm s}^{\rm c}  $
  ($ c_{\rm s}^{\rm c} \equiv \langle \hat{I}_{\rm s}^{\rm c} \rangle/ \langle \hat{I}_{\rm s} \rangle $)
  and (b) the pump chaotic component $ c_{\rm p}^{\rm ch} $
  ($ c_{\rm p}^{\rm ch} \equiv \langle \hat{I}_{\rm p}^{\rm ch} \rangle/ \langle \hat{I}_{\rm p} \rangle $)
  as they depend on pump power $ P $ for $ \gamma = 1 $ ($ \diamond $), $ \gamma = 0.5 $ ($ \ast
  $), $ \gamma = 0.1 $ ($ \triangle $), and $ \gamma = 0 $ ($ \circ $).}
\end{figure}

On the other hand, the originally coherent pump field attains also
its chaotic component due to its interaction with the signal and
idler fields being initially in the vacuum state [see Fig.~9(b)].
However, as the pump field as a whole is only weakly depleted, it
roughly preserves its initial coherent character. As a rule of
thumb, the number of coherent signal photons is roughly comparable
to the number of chaotic pump photons [compare the curves in
Figs.~8(b), 9(a) and 9(b)]. The pump field looses its photons
mainly in the central part of its spectrum, as shown in Fig.~10.
The spectral curve quantifying this effect closely reflects the
phase matching conditions at low pump powers $ P $ (see the curves
for $ P=1\times 10^{-8} $~W in Fig.~10). For greater pump powers $
P $, the appropriate curve is smoothed due to the complex exchange
of energy between the pump and the down-converted fields (see the
curves for $ P= 170 $~mW in Fig.~10). The comparison of curves in
Fig.~10 reveals that the presence of coherent components only
weakly influences the spectral curves.
\begin{figure}         
 \resizebox{0.8\hsize}{!}{\includegraphics{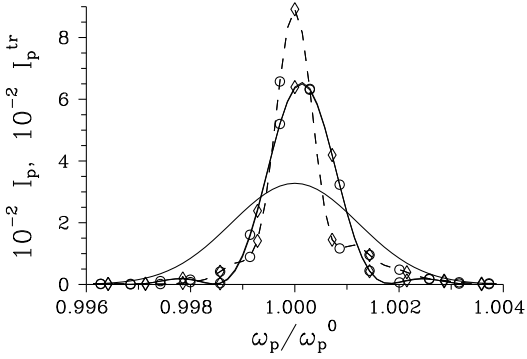}}
 \caption{Pump intensity spectral profile $ I_{\rm p} $ (solid curve without symbols) and profile $ I_{\rm p}^{\rm tr} $ of the pump intensity transferred to
  the TWB for $ \gamma = 0 $ ($ \circ $) and $ \gamma = 1 $ ($ \diamond $) for $ P = 1\times 10^{-8} $~W (nearly coinciding dashed curves) and
  $ P = 170 $~mW (nearly coinciding solid curves). The profiles are normalized such that $ \int d\omega_{\rm p} I_{\rm p}(\omega_{\rm p})/\omega_{\rm p}^0 = 1
  $ and $ \int d\omega_{\rm p} I_{\rm p}^{\rm tr} (\omega_{\rm p})/\omega_{\rm p}^0 = 1 $.}
\end{figure}

Also the number $ K $ of modes constituting the TWB is only weakly
affected by the coherent components (see Fig.~11).
\begin{figure}         
 \resizebox{0.8\hsize}{!}{\includegraphics{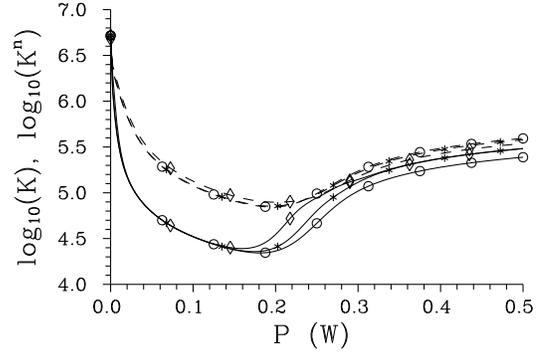}}
 \caption{Number of modes in the TWB determined by Eq.~(\ref{51}) ($ K^{\rm n} $, dashed curves)
  and Eq.~(\ref{52}) ($ K $, solid curves)
  as they depend on pump power $ P $ for $ \gamma = 1 $ ($ \diamond $), $ \gamma = 0.5 $ ($ \ast
  $) and $ \gamma = 0 $ ($ \circ $).}
\end{figure}
As shown in Fig.~11, the number $ K $ of modes decreases with the
increasing pump power $ P $ first and then it increases owing to
pump depletion inside the individual modes' triplets
\cite{PerinaJr2016a}. Numbers $ K $ and $ K^{\rm n} $ of modes
determined from photon-pair amplitude correlation functions and
photon-number statistics, respectively, are also compared in
Fig.~11: They both are suitable for the quantification of the
number of TWB modes. Whereas the first number is more suitable for
theoretical calculations, the second one is experimentally
available via the measurement of the photocount statistics
\cite{PerinaJr2013a}.

Spectral coherence of the TWB described by both the signal-field
intensity-fluctuation auto- and cross-correlation functions $
A_{{\rm s},\omega} $ and $ C_{\omega} $ behaves in the opposed way
to the number $ K $ of modes: It increases with the increasing
pump power $ P $ until its maximum ie reached at the threshold
pump power $ P_{\rm th} $ and then it decreases (see Fig.~12).
\begin{figure}         
 \resizebox{0.8\hsize}{!}{\includegraphics{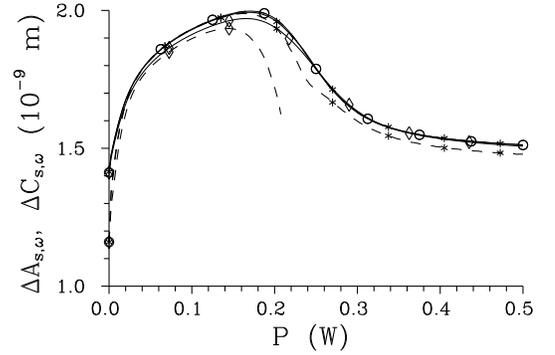}}
 \caption{Width $ \Delta A_{{\rm s},\omega} $ [$ \Delta C_{{\rm s},\omega} $]
  of intensity-fluctuation signal auto- [cross-] correlation function (solid [dashed] curves, FWHM)
  as it depends on pump power $ P $ for $ \gamma = 1 $ ($ \diamond $), $ \gamma = 0.5 $ ($ \ast
  $) and $ \gamma = 0 $ ($ \circ $). The curves giving $ \Delta A_{{\rm s},\omega} $ and $ \Delta C_{{\rm s},\omega} $ for
  $ \gamma = 0 $ nearly coincide; $ C_{{\rm s},\omega}(\omega_{\rm s}) \equiv C_{\omega}(\omega_{\rm s},\omega_{\rm i}^0) $.}
\end{figure}
Whereas the spectral profiles of intensity-fluctuation
auto-correlation functions $ A_{{\rm s},\omega} $ depend only
weakly on the strength of the coherent components (given by
parameter $ \gamma $), the spectral profiles of
intensity-fluctuation cross-correlation functions $ C_{\omega} $
remain practically unchanged for $ \gamma \in \langle 0,0.5
\rangle $. Positive correlations between the signal- and
idler-field intensity fluctuations in the vicinity of the central
peak and negative correlations outside this region are typical for
$ \gamma \in (0.5,1 \rangle $ (see Fig.~13 for the profiles of $
C_{\omega}^{\rm r} $ plotted for $ \gamma = 0.5 $, 0.8, and 1).
\begin{figure}         
 \resizebox{0.8\hsize}{!}{\includegraphics{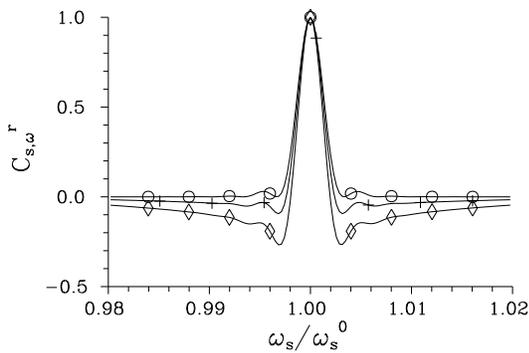}}
 \caption{Relative spectral intensity-fluctuation cross-correlation function $ C_{{\rm s},\omega}^{\rm r}(\omega_{\rm s}) $
  as it depends on relative signal frequency $ \omega_{\rm s}/\omega_{\rm s}^0 $ for $ \gamma = 1 $ ($ \diamond $), $ \gamma = 0.8 $
  (+) and $ \gamma = 0 $ ($ \circ $); $ C_{{\rm s},\omega}^{\rm r}(\omega_{\rm s}) \equiv C_{\omega}(\omega_{\rm s},\omega_{\rm i}^0) /
  {\rm max} (C_{\omega}(\omega_{\rm s},\omega_{\rm i}^0)) $, function $ {\rm max} $ gives the maximum, $ P = 170 $~mW.}
\end{figure}
This behavior is caused by the presence of intense signal and
idler coherent components whose generation is accompanied by
negative cross-correlations of intensity fluctuations (see Fig.~6
for individual modes' triplets and Fig.~15 below for the whole
fields). Width of the central peak in the intensity-fluctuation
cross-correlation function $ C_{\omega} $ then naturally decreases
with the increasing values of parameter $ \gamma $ (see the curve
in Fig.~13 for $ \gamma = 1 $). For the pump powers $ P $ greater
than 200~mW and $ \gamma = 1 $, the central peak in the
intensity-fluctuation cross-correlation function is even lost and
it is replaced by a dip. Hand in hand, a shallow dip is formed in
the signal (and idler) intensity spectral profiles. We note that
the widths of the signal (and idler) intensity spectral profiles
decrease with the increasing pump power $ P $ until $ P_{\rm th} $
is reached and then they increase. This is another manifestation
of the internal dynamics of TWBs. Detailed relationship between
the internal dynamics of a TWB and its properties is discussed in
\cite{PerinaJr2016}.

The most striking feature of ideal (i.e. noiseless) chaotic TWBs
is their perfect signal-idler intensity sub-shot-noise
cross-correlation quantified by parameter $ \tilde{R}_{\rm si} $
defined in Eq.~(\ref{48}) ($ \tilde{R}_{\rm si}=0 $). This
originates in the ideal pairwise character of chaotic TWBs in
which each signal photon is accompanied by its own idler twin. The
coherent components in both the signal and idler fields disturb
this ideal pairing. This results in the gradual suppression of
sub-shot-noise intensity correlations, which has been reported in
the experiment \cite{Jedrkiewicz2004}. As shown in Fig.~14, the
greater the parameter $ \gamma $ the greater the values of
parameter $ \tilde{R}_{\rm si} $.
\begin{figure}         
 \resizebox{0.8\hsize}{!}{\includegraphics{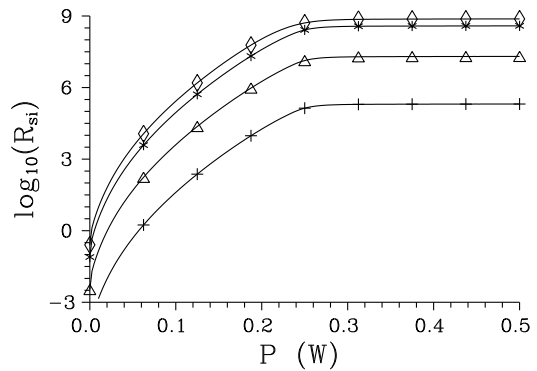}}
 \caption{Parameter $ \tilde{R}_{\rm si} $ determining the signal-idler sub-shot-noise
  intensity correlations as it depends on pump power $ P
  $ for $ \gamma = 1 $ ($ \diamond $), $ \gamma = 0.5 $ ($ \ast
  $), $ \gamma = 0.1 $ ($ \triangle $), and $ \gamma = 0.01 $ (+).}
\end{figure}
Also, the greater the pump power $ P $ the greater the values of
parameter $ \tilde{R}_{\rm si} $. The intensity sub-shot-noise
correlations are completely lost for greater pump powers $ P $.
However, classical correlations between the signal- and
idler-field intensities remain also for greater pump powers $ P $
[for the reduced signal-idler intensity-fluctuation moment $
\tilde{r}_{\rm si} $ defined in Eq.~(\ref{50}), see Fig.~15].
\begin{figure}         
 \resizebox{0.8\hsize}{!}{\includegraphics{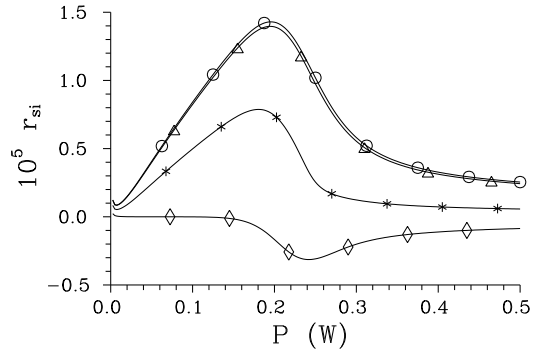}}
 \caption{Reduced signal-idler intensity-fluctuation moment $ \tilde{r}_{\rm si} $ as it depends on pump power $ P
  $ for $ \gamma = 1 $ ($ \diamond $), $ \gamma = 0.5 $ ($ \ast
  $), $ \gamma = 0.1 $ ($ \triangle $), and $ \gamma = 0 $ ($ \circ
  $).}
\end{figure}
The signal-idler intensity-fluctuation correlations are positive
for $ \gamma \in \langle 0,0.5 \rangle $ and attain their maximal
values for pump powers around $ P_{\rm th} $. However, considering
$ \gamma \in (0.5,1 \rangle $, the correlations are negative for
greater pump powers $ P $ as a consequence of the presence of more
intense coherent components \cite{Mista1969,Perinova1981}. We note
that the overall signal-field photon-number statistics approach
the Poissonian distributions as the signal field is composed of
many modes ($ \tilde{r}_{\rm s} = 1 $).

\section{Interference patterns in sum-frequency generation and Hong-Ou-Mandel interferometer}

In this section, we discuss the influence of coherent components
on the temporal correlation functions arising in the mutual
interference of the signal and idler fields.

First, we analyze the intensity $ I^{\rm SFG} $ of the
sum-frequency field generated by the mutually delayed signal and
idler fields. The profile of intensity $ I^{\rm SFG} $ plotted
versus the time delay $ \tau $ is composed of two superimposed
peaks (see Fig.~16).
\begin{figure}         
 \resizebox{.45\hsize}{!}{\includegraphics{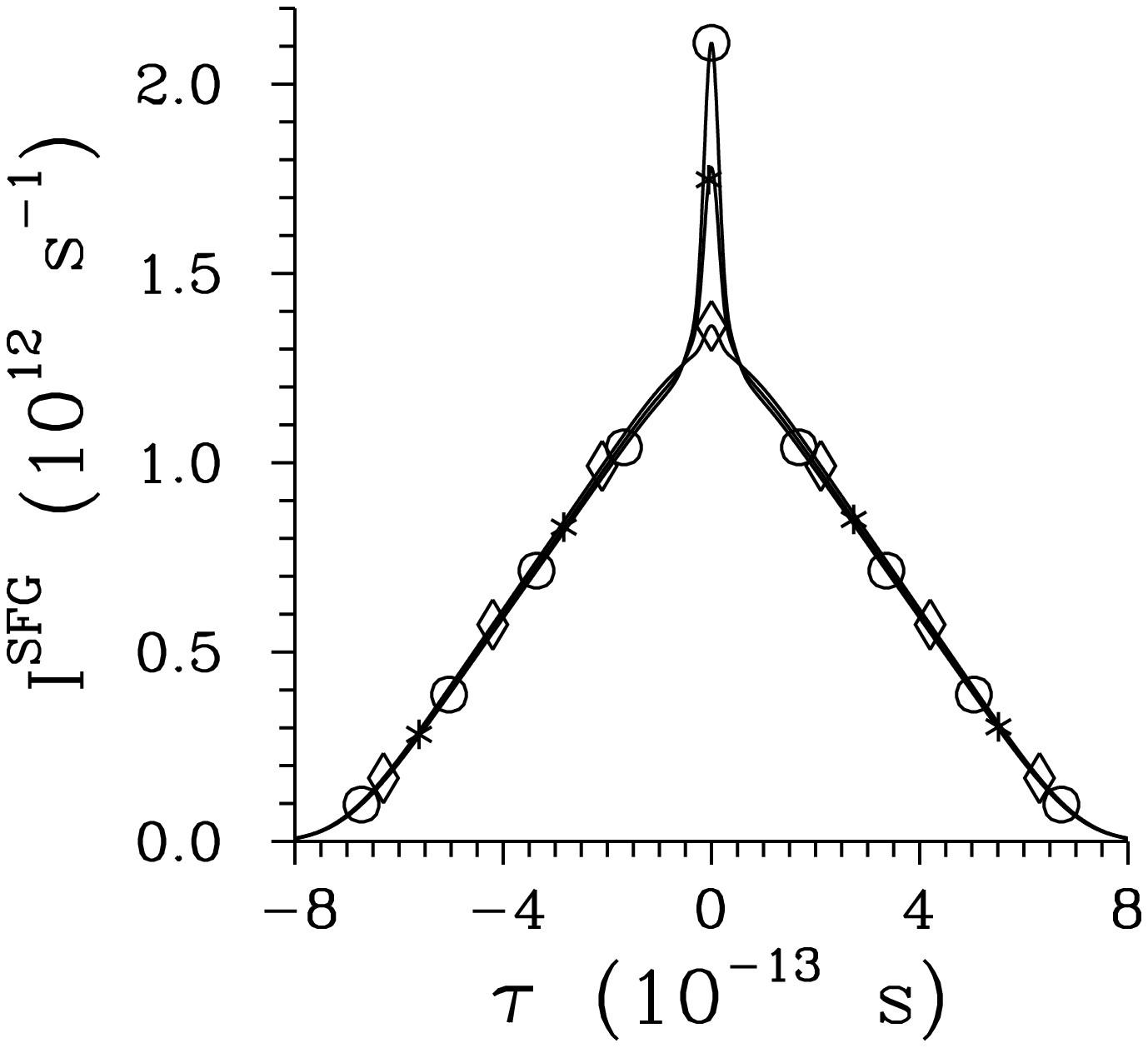}}
  \hspace{2mm}
 \resizebox{.45\hsize}{!}{\includegraphics{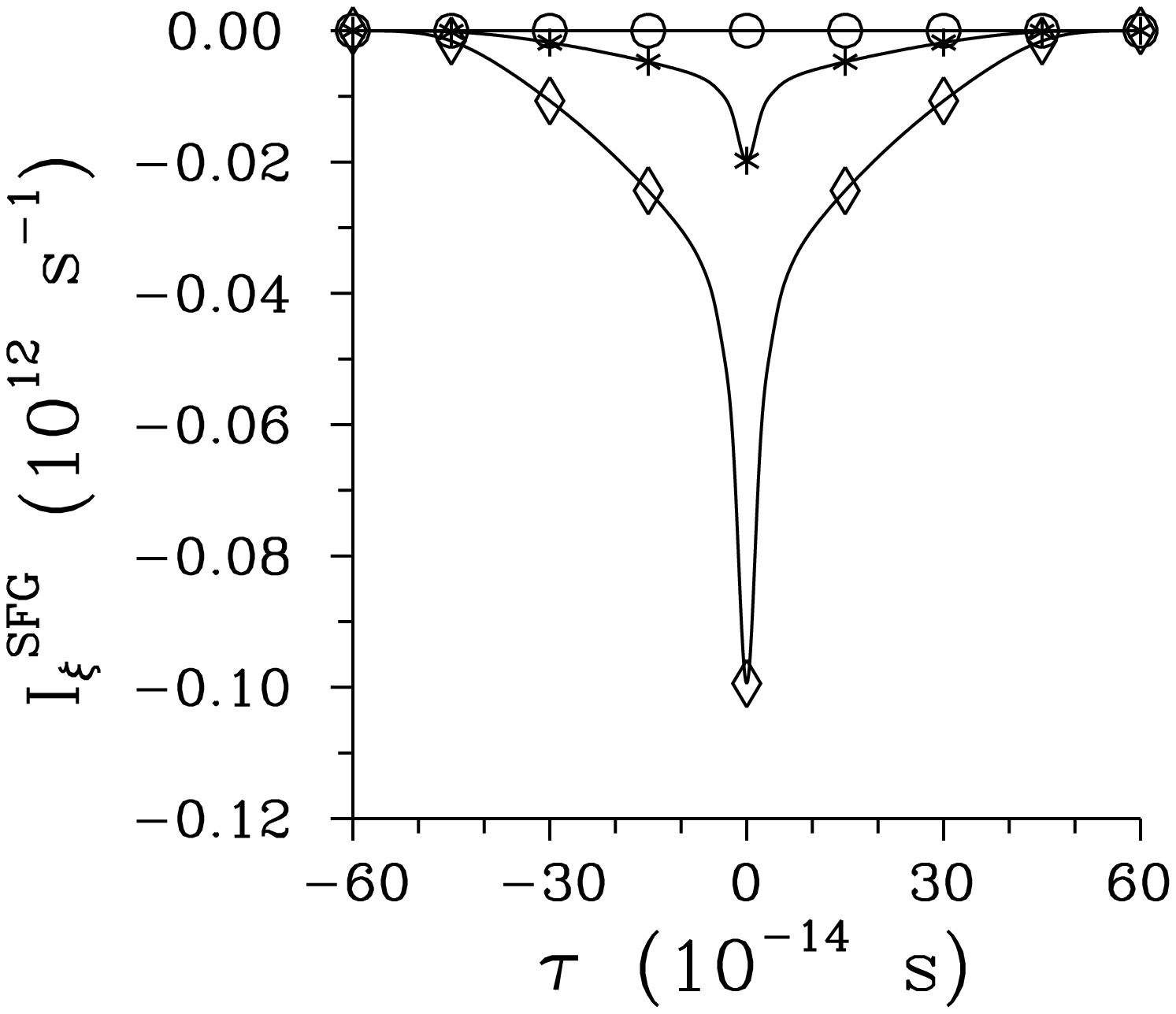}}

 \centerline{ (a) \hspace{.4\hsize} (b) }
 \caption{(a) Intensity $ I^{\rm SFG} $ of the sum-frequency field and
  (b) its contribution $ I^{\rm SFG}_{\xi} $ created by the
  coherent components [the fourth term in Eq.~(\ref{64})]
  as functions of time delay $ \tau $ for $ \gamma = 1 $ ($ \diamond $), $ \gamma = 0.5 $ ($ \ast
  $) and $ \gamma = 0 $ ($ \circ $); $ P=170 $~mW. The field is normalized such that
  $ \int d\tau I^{\rm SFG}(\tau) = 1 $.}
\end{figure}
A broad peak is formed by the temporal intensity profiles of the
'independent' signal and idler fields [see the first term in
Eq.~({\ref{64})]. When the coherent components occur, they modify
the broad peak mainly in its central part, where they slightly
reduce the intensities [see the fourth term in Eq.~(\ref{64})]. On
the other hand, a narrow peak arises due to the amplitude
cross-correlations between the signal and idler fields. Roughly
speaking, it is formed by the interference of the photons
belonging to the same photon pair. As such, the narrow peak's
height decreases with the increasing influence of the coherent
components. The reason is that the coherent components lower the
number of paired photons [via reducing coefficients $ d $ in the
second term in Eq.~({\ref{64})]. On the other hand, the coherent
components create 'linear coupling' between the signal and idler
fields which, however, has only negligible influence on the narrow
peak [through the third term in Eq.~({\ref{64})].

These mechanisms explain the behavior of widths $ \Delta I^{\rm
SFG}_{\rm b} $ and $ \Delta I^{\rm SFG}_{\rm n} $ of the broad and
narrow peaks, respectively, as the pump power $ P $ varies.
Whereas the width $ \Delta I^{\rm SFG}_{\rm b} $ of the broad peak
narrows for the pump powers $ P $ around the threshold power $
P_{\rm th} $ [see Fig.~17(a)], the width $ \Delta I^{\rm SFG}_{\rm
n} $ of the narrow peak broadens in this area [see Fig.~17(b)].
\begin{figure}         
 \resizebox{.45\hsize}{!}{\includegraphics{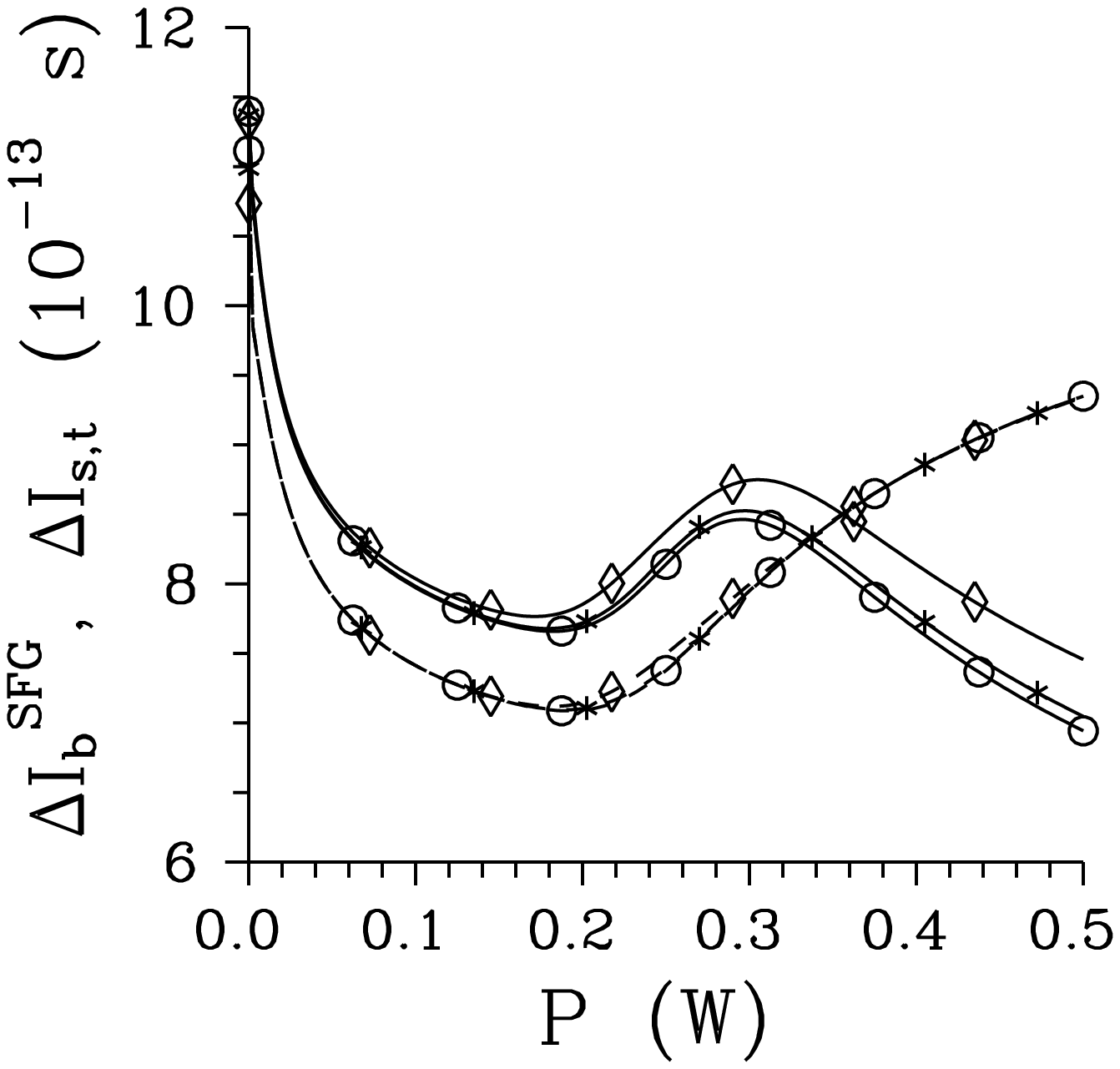}}
  \hspace{2mm}
 \resizebox{.45\hsize}{!}{\includegraphics{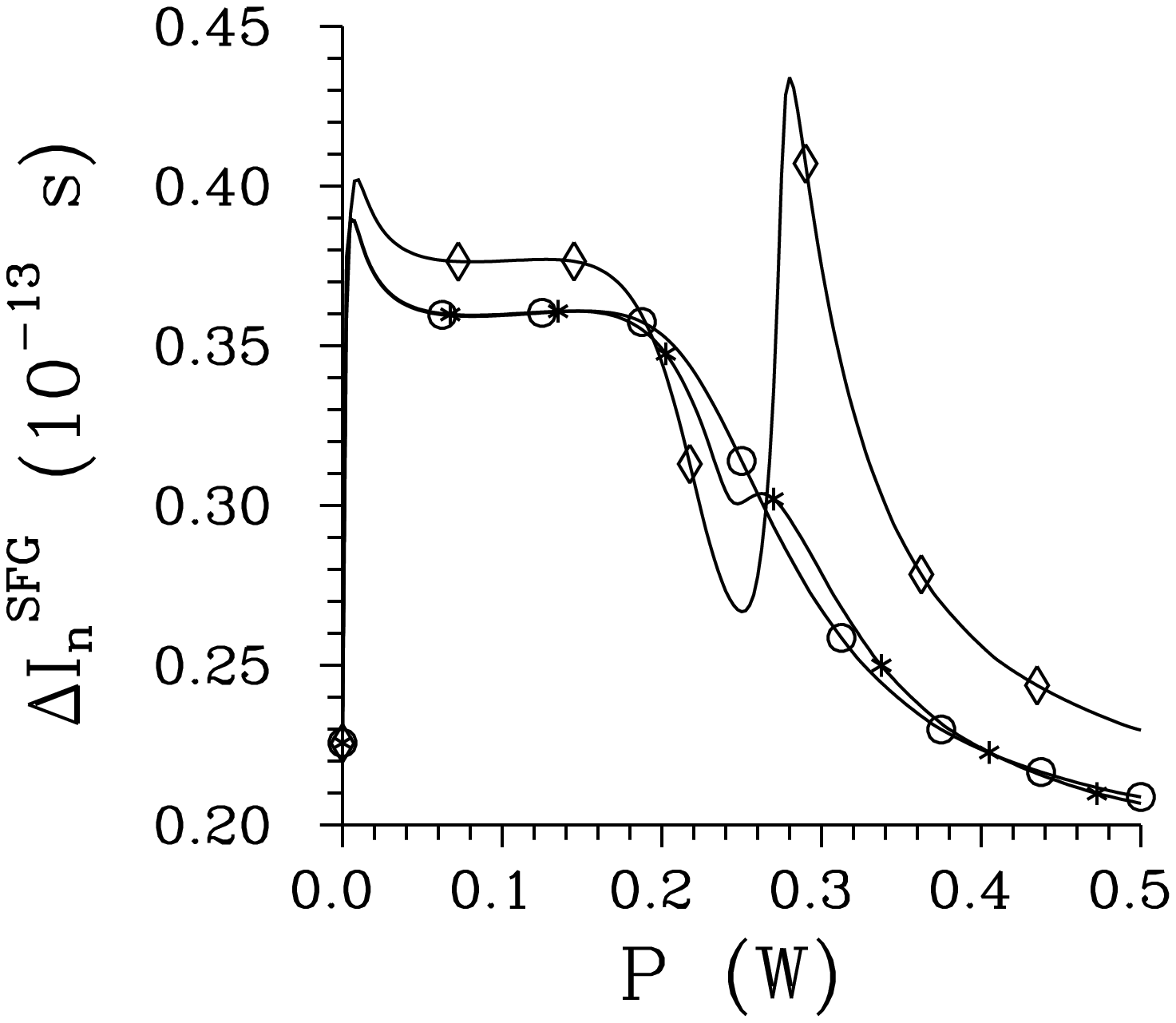}}

 \centerline{ (a) \hspace{.4\hsize} (b) }
 \caption{(a) Width $ \Delta I^{\rm SFG}_{\rm b} $ of the broad intensity peak of
  the sum-frequency field (solid curves, FWHM) and width $ \Delta I_{{\rm s},t} $ of the signal-field
  photon flux (dashed curves, FWHM) and (b) width $ \Delta I^{\rm SFG}_{\rm n} $ of the narrow intensity peak of
  the sum-frequency field (FWHM) as they depend on pump power $ P $ for $ \gamma = 1 $ ($ \diamond $), $ \gamma = 0.5 $ ($ \ast
  $) and $ \gamma = 0 $ ($ \circ $).}
\end{figure}
This corresponds to the narrowing of the temporal intensity
profiles of the signal and idler fields and broadening of the
temporal intensity cross-correlation functions. The increase of
width $ \Delta I^{\rm SFG}_{\rm n} $ of the narrow peak for the
pump powers $ P $ above the threshold power $ P_{\rm th} $
observed in the curve drawn for $ \gamma = 1 $ in Fig.~17(b)
indicates the dominance of the coherent components above the
chaotic contributions (for details, see the discussion of the
Hong-Ou-Mandel interference below).

Considering only chaotic TWBs the visibility $ V^{\rm SFG} $ of
the narrow peak depends only weakly on pump power $ P $ (see
Fig.~18). The coherent components considerably reduce the
visibility $ V^{\rm SFG} $, up to 1/7 for pump powers $ P \leq
P_{\rm th} $ and even 1/20 for $ P > P_{\rm th} $ for the analyzed
case and $ \gamma = 1 $.
\begin{figure}         
 \resizebox{0.8\hsize}{!}{\includegraphics{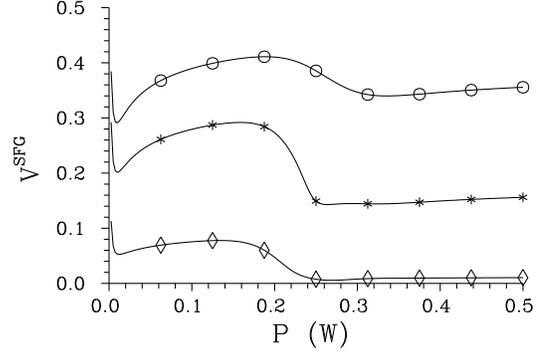}}
 \caption{Visibility $ V^{\rm SFG} $ of the narrow intensity peak of the sum-frequency field
  as it depends on pump power $ P $ for $ \gamma = 1 $ ($ \diamond $), $ \gamma = 0.5 $ ($ \ast
  $) and $ \gamma = 0 $ ($ \circ $); $ V^{\rm SFG} \equiv I^{\rm SFG}_{\rm n}/( I^{\rm SFG}_{\rm n} +
  I^{\rm SFG}_{\rm b}) $ where $ I^{\rm SFG}_{\rm n} $ ($ I^{\rm SFG}_{\rm b} $) gives the intensity
  of the narrow (broad) peak.}
\end{figure}

Two different time constants clearly visible in the intensity
profile $ I^{\rm SFG}(\tau) $ of the sum-frequency field are also
observed in the normalized intensity-fluctuations correlation
function $ R_{\rm n}^{\Delta} $ plotted as a function of the
mutual time delay $ \tau $ between the signal and idler fields
propagating in the Hong-Ou-Mandel interferometer (see Fig.~19).
\begin{figure}         
 \resizebox{.45\hsize}{!}{\includegraphics{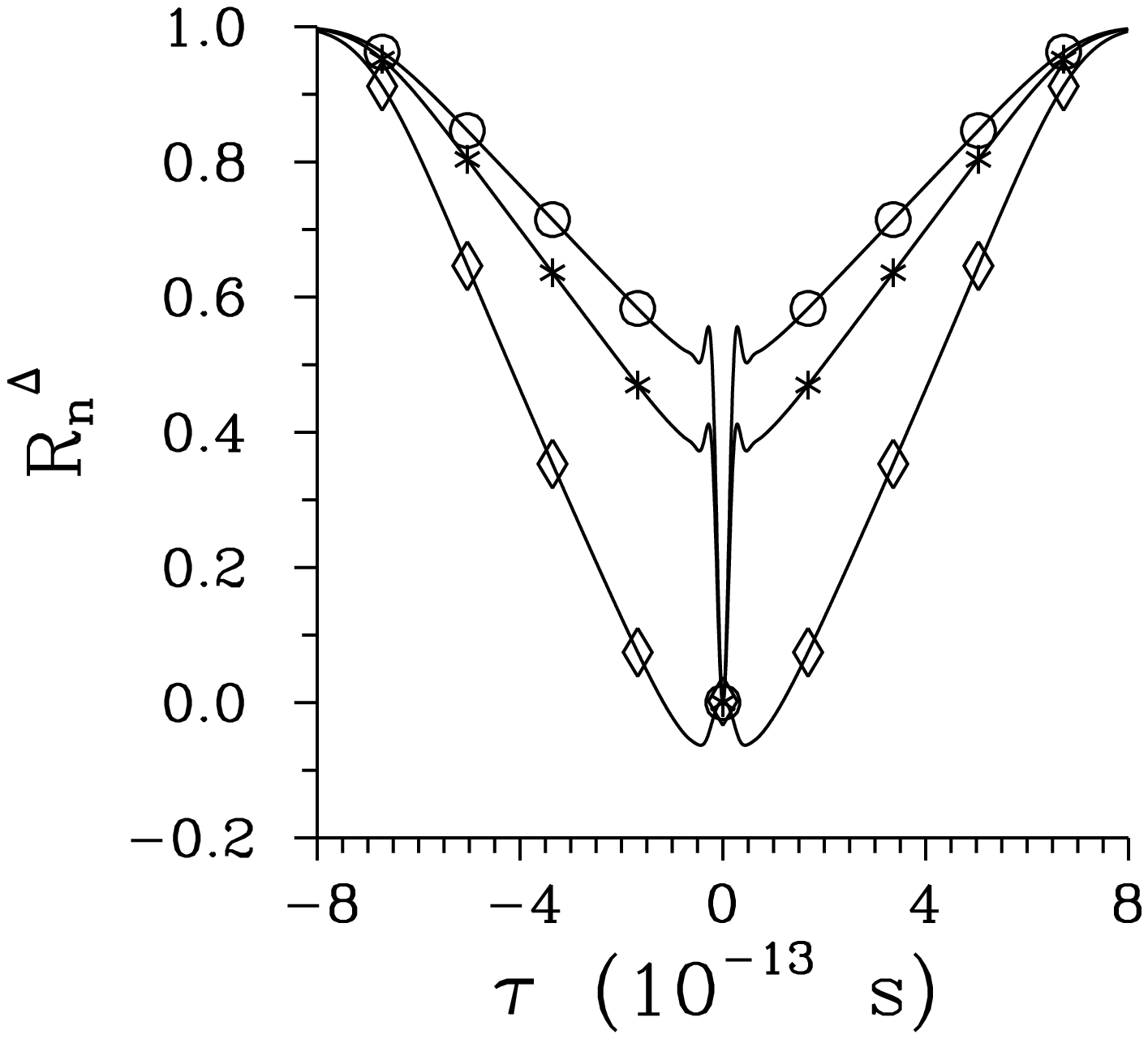}}
  \hspace{2mm}
 \resizebox{.45\hsize}{!}{\includegraphics{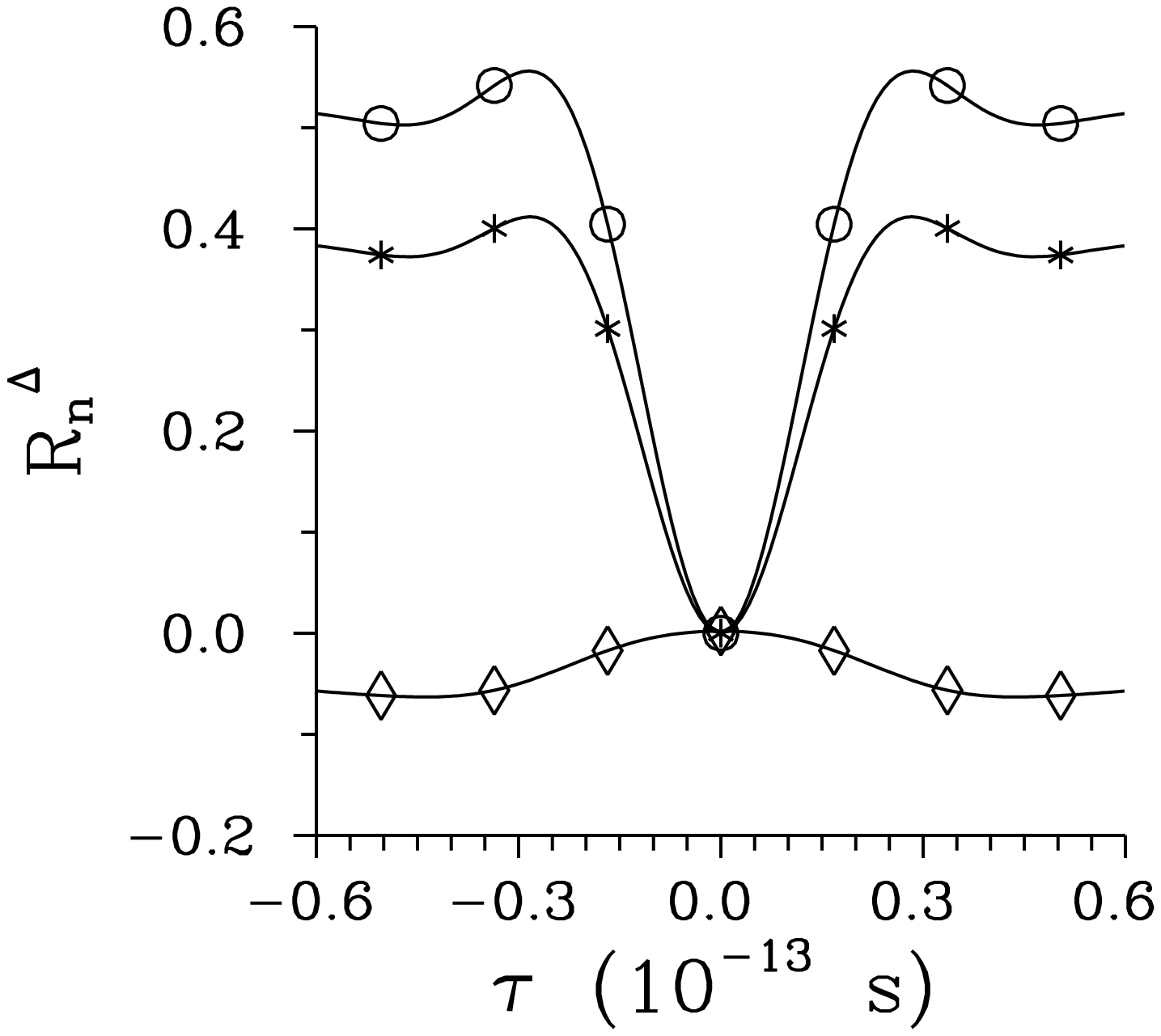}}

 \centerline{ (a) \hspace{.4\hsize} (b) }
 \caption{(a) Normalized intensity-fluctuations correlation function $ R^{\Delta}_{\rm n} $
  in the Hong-Ou-Mandel interferometer as a function of time delay $ \tau $ for $ \gamma = 1 $ ($ \diamond $), $ \gamma = 0.5 $ ($ \ast
  $) and $ \gamma = 0 $ ($ \circ $); $ P = 170 $~mW. In (b), detail of the central part of $ R^{\Delta}_{\rm n}(\tau) $ is plotted.}
\end{figure}
For chaotic TWBs, the profile $ R_{\rm n}^{\Delta} $ consists of
two central dips. Similarly as in the case of sum-frequency
generation, the broader dip is related to the mutual overlap of
the signal- and idler-field intensities, whereas the narrower dip
arises from intensity cross-correlations of the fields. The
presence of the coherent components qualitatively changes the
profile of correlation function $ R_{\rm n}^{\Delta} $, as their
contribution to the correlation function $ R_{\rm n}^{\Delta} $
differs in the sign compared to that of the chaotic contribution.
As a consequence, the original narrow dip in a chaotic TWB
gradually changes into a narrow peak when the role of coherent
components increases. As the coherent components influence also
the asymptotic values of correlation function $
R^{\Delta}(\tau\rightarrow \pm \infty) $ used in the
normalization, we even observe negative values of the normalized
correlation function $ R_{\rm n}^{\Delta} $ (see the curve for $
\gamma=1 $ in Fig.~19). It holds that the greater the parameter $
\gamma $, the larger the coherent contributions and also the lower
the chaotic contribution to the correlation function $ R_{\rm
n}^{\Delta} $.

Visibility $ V^{\rm HOM} $ of the normalized
intensity-fluctuations correlation function $ R_{\rm n}^{\Delta} $
remains maximal ($  V^{\rm HOM} = 1 $) with the increasing pump
power $ P $ only for purely chaotic TWBs [see Fig.~20(a)]. TWBs
with nonzero coherent components partly loose their visibility in
the area around the threshold power $ P_{\rm th} $. Closer
inspection of visibilities $ V^{\rm  HOM}_{\xi} $ and $ V^{\rm
HOM}_d $ characterizing the coherent and chaotic contributions to
the correlation function $ R_{\rm n}^{\Delta} $, respectively,
reveals qualitatively different behavior of both contributions. We
note that the coherent (chaotic) contribution is described by the
first [second] term in Eq.~(\ref{72}). Whereas the visibility $
V^{\rm HOM}_d $ of the chaotic contribution drops down with the
increasing pump power $ P $, the visibility $ V^{\rm HOM}_\xi $ of
the coherent contribution increases [see Fig.~20(b)]. Both
contributions are balanced for pump powers $ P \leq P_{\rm th} $
and so the overall visibility  $ V^{\rm HOM} $ remains close to
one in this region. However, this balance is lost for greater pump
powers $ P > P_{\rm th} $ which results in lower values of the
visibility $ V^{\rm HOM} $ observed in this region.
\begin{figure}         
 \resizebox{.45\hsize}{!}{\includegraphics{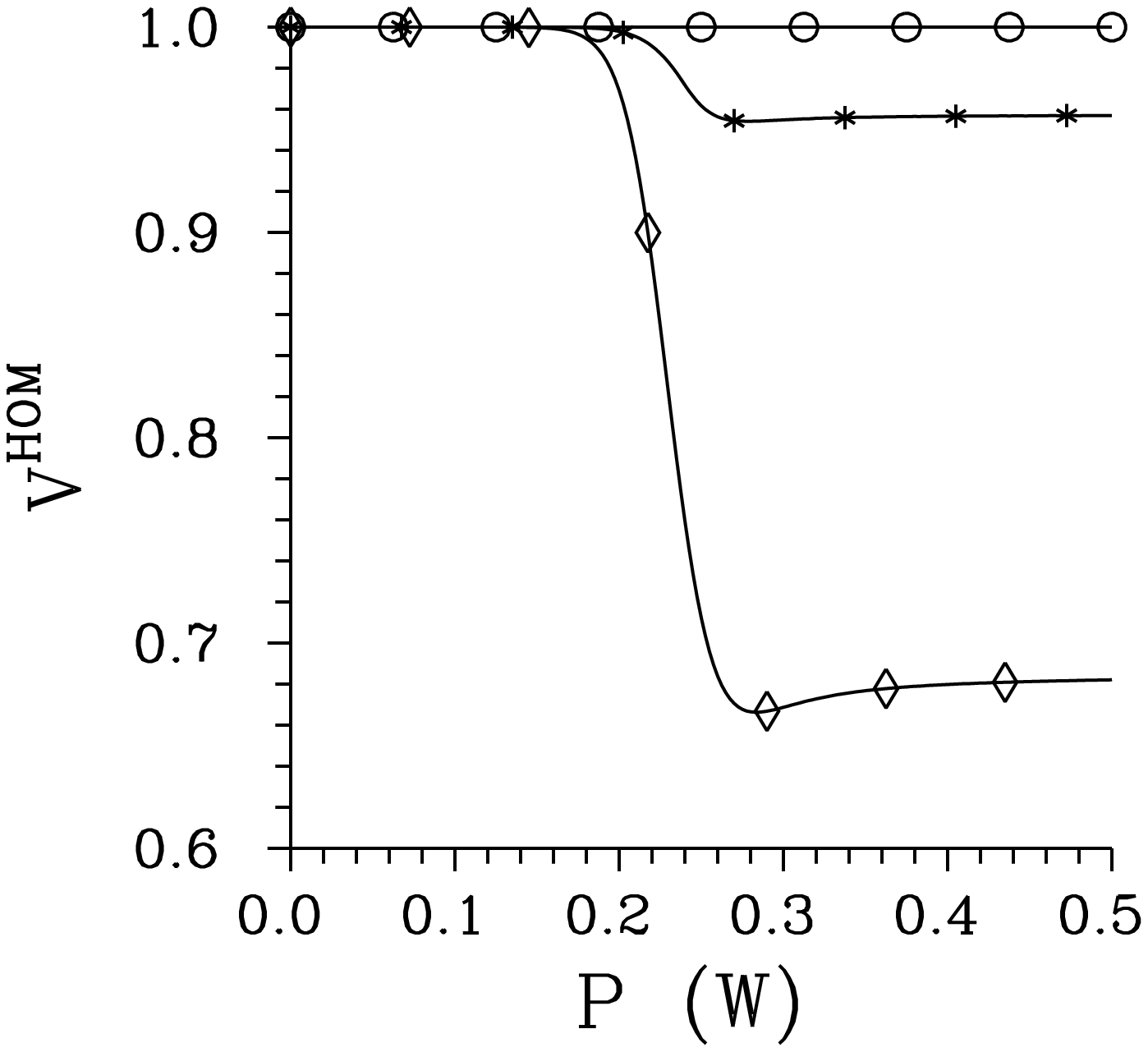}}
  \hspace{2mm}
 \resizebox{.45\hsize}{!}{\includegraphics{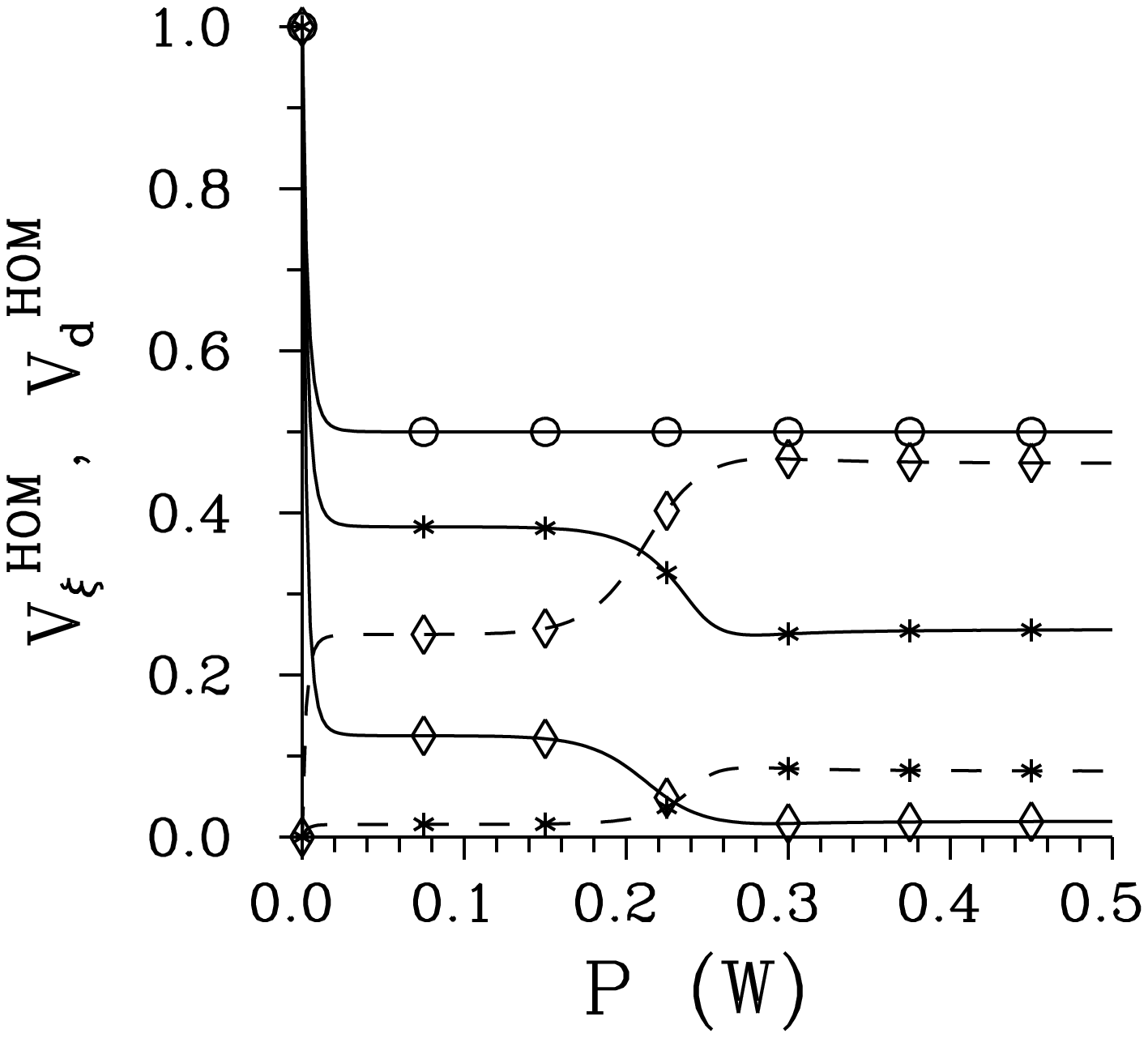}}

 \centerline{ (a) \hspace{.4\hsize} (b) }
 \caption{(a) Visibility $ V^{\rm HOM} $ of the normalized
  intensity-fluctuations correlation function $
  R_{\rm n}^{\Delta} $ as
  a function of pump power $ P $ for $ \gamma = 1 $ ($ \diamond $), $ \gamma = 0.5 $ ($ \ast
  $) and $ \gamma = 0 $ ($ \circ $). In (b) visibility  $ V^{\rm
  HOM}_{\xi} $ [$ V^{\rm HOM}_d $] of the coherent [chaotic]
  contribution as given by the first [second] term in Eq.~(\ref{72})
  is plotted (dashed [solid] curves); $ V^{\rm HOM} = (R_{{\rm n},{\rm max}}^\Delta - R_{{\rm n},{\rm min}}^\Delta)/
  R_{{\rm n},{\rm max}}^\Delta $, where $ R_{{\rm n},{\rm max}}^\Delta $
  ($ R_{{\rm n},{\rm min}}^\Delta $) denotes the maximal (minimal)
  value.}
\end{figure}
We note that the visibility of the normalized intensity
correlation function $ R_{\rm n} $ defined in Eq.~(\ref{70}) is
practically negligible for the discussed pump powers $ P $. This
visibility equals one for low pump powers $ P $
\cite{PerinaJr1999a} and then it drops fast close to zero when the
pump power $ P $ increases \cite{Iskhakov2013,Cosme2008}.

Also the widths of the dips and peaks in the correlation functions
change with the increasing pump power $ P $. The width $ \Delta
R^{\Delta}_{\rm b} $ of the broad dip in the correlation function
$ R_{\rm n}^{\Delta} $ attains its (local) minimum in the area of
pump powers around $ P_{\rm th} $ [see Fig.~21(a)]. Contrary to
this, the widths $ \Delta R^{\Delta}_{{\rm n},\xi} $ and $ \Delta
R^{\Delta}_{{\rm n},d} $ of the narrow coherent peak and narrow
chaotic dip, respectively, reach their maxima in this area [see
Fig.~21(b)]. As it is evident from the curves plotted in
Fig.~21(b) the coherent peak is broader than the chaotic dip, but
both together form a narrow central dip or peak. The curves in
Fig.~21 show that the coherent components influence only weakly
the characteristic temporal widths of the correlation function $
R_{\rm n}^{\Delta} $.
\begin{figure}         
 \resizebox{.45\hsize}{!}{\includegraphics{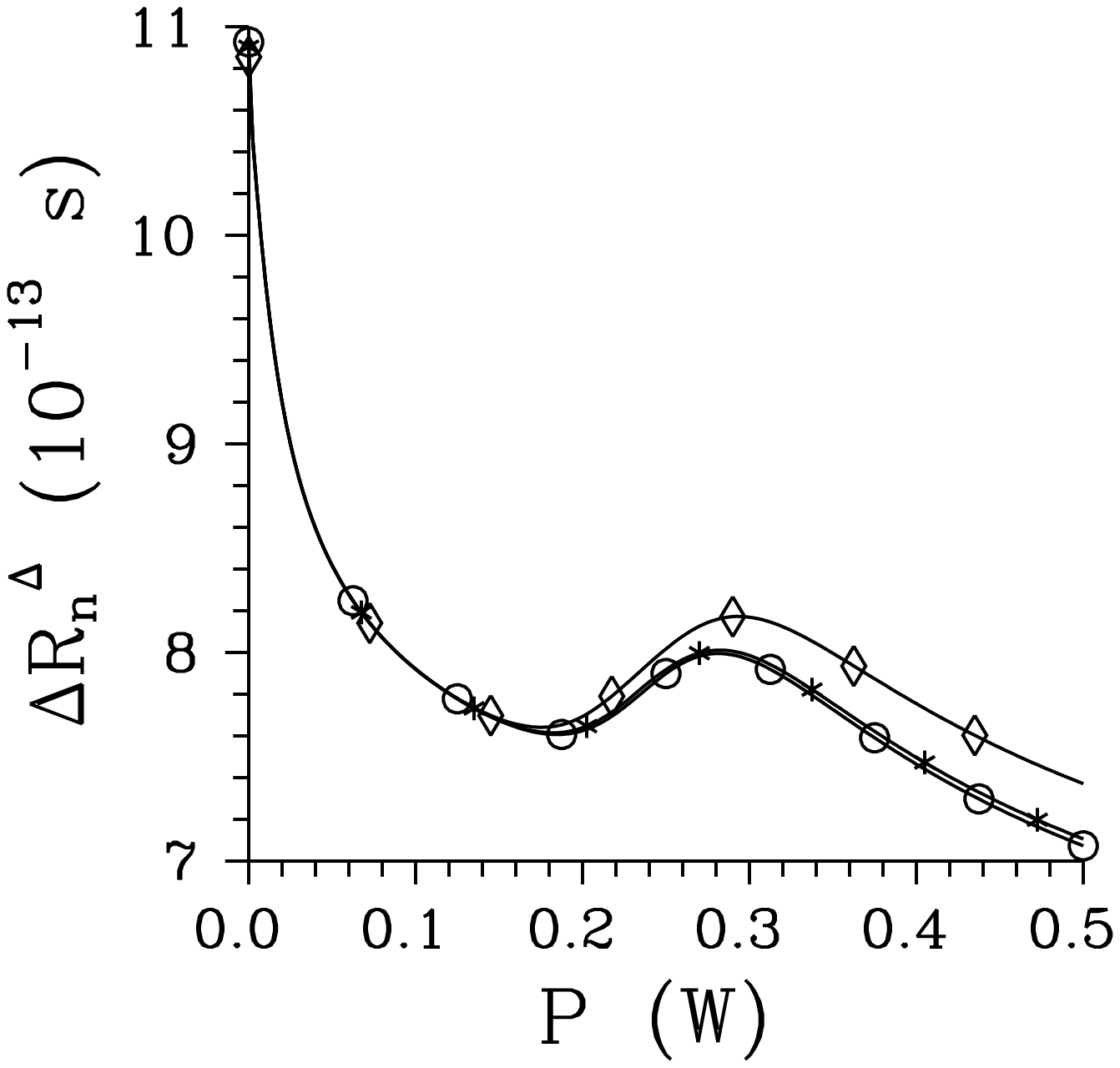}}
  \hspace{2mm}
 \resizebox{.45\hsize}{!}{\includegraphics{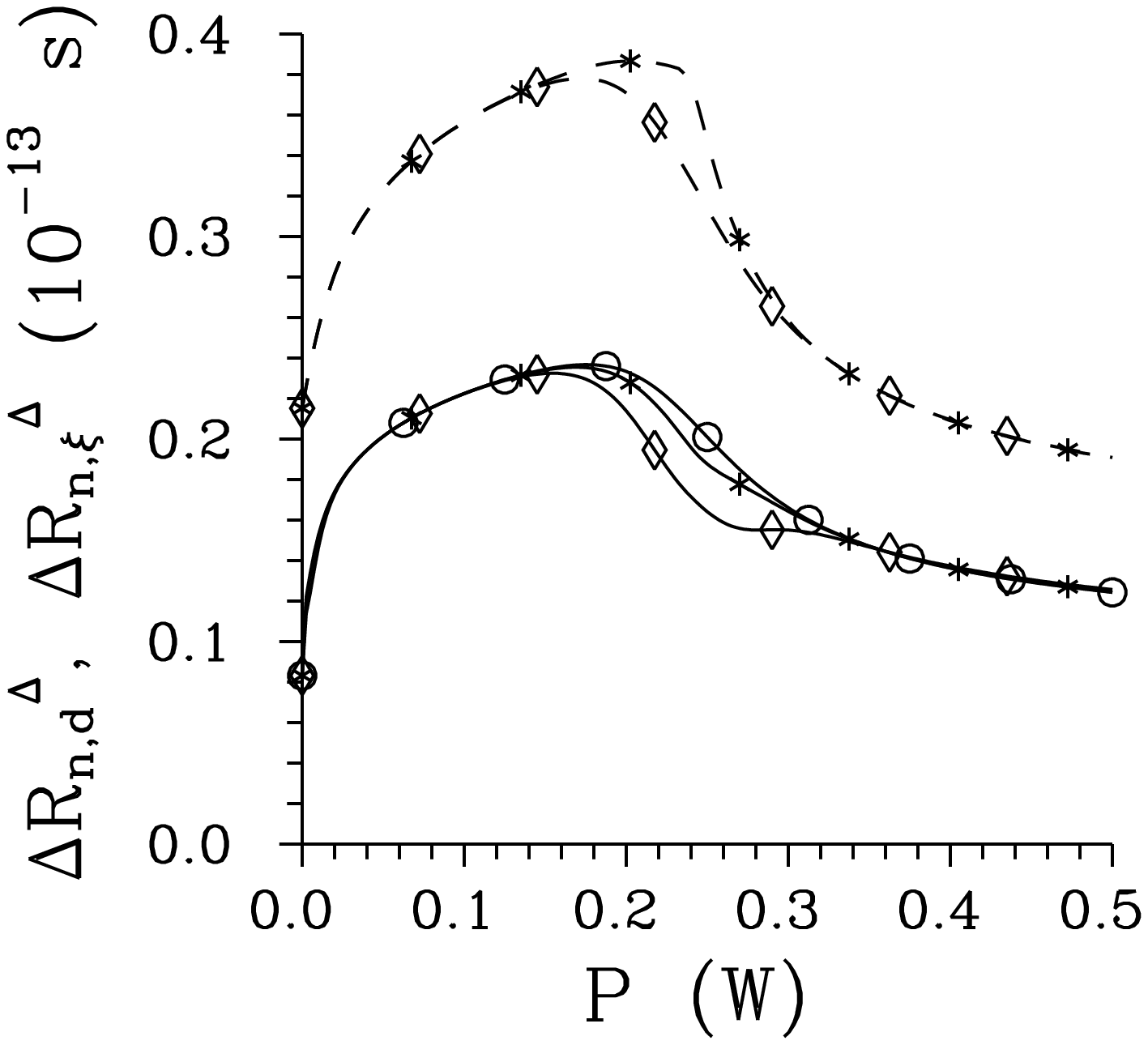}}

 \centerline{ (a) \hspace{.4\hsize} (b) }
 \caption{(a) Width $ \Delta R^{\Delta}_{\rm n} $ of the broad dip in the normalized intensity-fluctuations correlation
  function $ R_{\rm n}^{\Delta} $ (FWHM) and
  (b) width $ \Delta R^{\Delta}_{{\rm n},\xi} $ [$ \Delta R^{\Delta}_{{\rm n},d} $] of the narrow
  peak [dip] caused by the coherent [chaotic] contribution (the first [second] term in
  Eq.~(\ref{72}), FWHM, dashed [solid] curves) as they depend on pump power $ P $;
  $ \gamma = 1 $ ($ \diamond $), $ \gamma = 0.5 $ ($ \ast
  $) and $ \gamma = 0 $ ($ \circ $).}
\end{figure}

\section{Conclusions}

Treating the nonlinear dynamics of intense parametric
down-conversion in terms of individual modes' triplets we were
able to investigate the nonlinear process in the regime with pump
depletion. The generation of coherent components in the otherwise
chaotic signal and idler fields has been revealed. The influence
of coherent components on the properties of the emitted intense
twin beams has been analyzed considering twin-beam intensity,
signal and idler intensity correlations, numbers of modes
constituting the twin beam, and spectral intensity auto- and
cross-correlation functions versus the pump power. The analysis of
the interference patterns observed in the process of sum-frequency
generation and in the Hong-Ou-Mandel interferometer has revealed
that both of them allow for the experimental observation of the
coherent components in intense twin beams. Whereas the coherent
components considerably reduce the narrow peak in the interference
pattern of sum-frequency generation, they form a narrow peak at
the bottom of a broad dip in the Hong-Ou-Mandel interference
pattern. We believe that the obtained theoretical results will
stimulate further experimental investigations of intense twin
beams.

\section*{Acknowledgments}
The author thanks A. Luk\v{s} for the discussion concerning the
analytical solution of Eqs.~(\ref{22}) and (\ref{23}). He also
acknowledges discussions with J. Pe\v{r}ina, M. Bondani,
O.~Haderka, and A. Allevi. Support from projects No.~15-08971S of
GA \v{C}R and No.~LO1305 of M\v{S}MT \v{C}R is acknowledged.


\end{document}